\ifdefined\paperonly
\documentclass[acmsmall,screen,review]{acmart}
\else
\documentclass[acmsmall,screen]{acmart}
\fi
\setcopyright{none}

\AtBeginDocument{%
  \providecommand\BibTeX{{%
    \normalfont B\kern-0.5em{\scshape i\kern-0.25em b}\kern-0.8em\TeX}}}

\usepackage{hyperref, wrapfig, amsthm, pict2e, booktabs, subcaption,
  subfiles, rotating, pifont, indentfirst, amsmath, amsfonts,
  mathrsfs, adjustbox, mathtools,booktabs, microtype, ebproof,
  enumitem, multirow, multicol, setspace, xcolor, lineno, listings,
  stmaryrd, xspace, stackengine, makecell}

\usepackage{empheq}

\usepackage[tableposition=top]{caption}
\usepackage{tikz-cd}
\usetikzlibrary{decorations.pathmorphing}
\usepackage[ruled,linesnumbered,vlined]{algorithm2e}

\SetCommentSty{mycommfont}
\usepackage{graphicx}
\usepackage[frozencache,cachedir=minted-cache]{minted}
\usepackage{tikz}
\newcounter{markeq}
\usepackage{spacingtricks}

\usepackage[T1]{fontenc}

\usepackage[xcolor]{changebar}
\cbcolor{blue}

\usepackage{spacingtricks, commands, refinementtydef}

\begin{document}

\title[Context Types]{Context Types: Bunched Modalities for Unifying Safety and Reachability}
\author{Zhe Zhou}
\authornote{Work performed while at Purdue University}
\affiliation{
  \institution{Peking University}            %
  \country{China}                    %
}

\author{Benjamin Delaware}
\affiliation{
  \institution{Purdue University}            %
  \country{USA}                    %
}

\author{Suresh Jagannathan}
\affiliation{
  \institution{Purdue University}            %
  \country{USA}                    %
}

\begin{abstract}
  Most type systems are designed to verify safety properties: the
  judgement $e : \tau$ holds if $e$ \emph{may} evaluate to any value
  contained in the denotation of $\tau$, but nothing more. More recent
  type systems have considered the complementary class of reachability
  properties: in these systems, the typing judgment holds if $e$
  \emph{must} be able to evaluate to every value denoted by $\tau$,
  and possibly more. This paper introduces a refinement type system
  for higher-order functional programs where types can capture both
  safety and reachability properties. Simultaneously supporting both
  is challenging because each imposes different constraints on how
  variable bindings in a typing context are interpreted.  We address
  this challenge by treating typing contexts as a kind of resource
  that mediates which capabilities are available when typing a
  program. We formalize this notion by developing a resource algebra
  of contextual capabilities and a bunched modal context logic over
  this algebra. We then interpret our type system in this logic,
  establishing soundness with respect to a denotational semantics that
  unifies safety and reachability guarantees within a single typing
  judgement.
 \end{abstract}

\maketitle

\section{Introduction} %
\label{sec:intro}

Static type systems are one of the most versatile and widely-used
program reasoning techniques available today. While their particulars
can vary wildly, common to all type systems are two key ingredients:
the use of \emph{types} to specify program behavior, and some
deductive mechanism for \emph{compositionally verifying} that the
program is consistent with that specification.  Type systems have
traditionally targeted \emph{safety} properties, a focus embodied by
Milner's famous slogan that ``well-typed programs cannot {`go
wrong'}''~\cite{Milner+78}. Here, the typing judgement $\Gamma \vdash e
: \tau$ asserts that $e$ \emph{may} evaluate to many values (under any
environment consistent with $\Gamma$), but that set of values is
entirely contained within the ``safe'' region specified by $\tau$.

Dually, \emph{reachability} properties seek to precisely capture a
subset of the values that a program \emph{must} reach. Rather than
ruling out certain executions, reachability analyses seek to certify
the existence of feasible executions.  Reachability analyses are a
well-established and practical technique for informing program
optimizations and transformations~\cite{WS94,GM21,JTW+98}.  They have
also been effectively used to find buggy behaviors~\cite{BMC+03}; for
example, several recent works~\cite{GOS19,ReverseHoareLogic, OHP19,
  ZDS23} have studied deductive ``incorrectness'' reasoning frameworks
which can prove the existence of witnesses of safety violations, with
a goal of avoiding the high false positive rate reported by typical
bug detection tools.  The question of how to develop type systems for
reachability, however, has been largely unexplored.  Recent
works~\cite{RefinementTypeRefutations, CoverageType, Ill-types} that
have studied type-based reachability reasoning have either focused on
a particular property, e.g., checking completeness of test input
generators~\cite{CoverageType}, or a particular application, e.g.,
generating concrete counterexamples refuting refinement type
specifications~\cite{RefinementTypeRefutations} or proving ill-typed
programs do not evaluate~\cite{Ill-types}, rather than providing a
general type-based reachability framework.

To unify these two views of types, this paper proposes a novel
semantic model that can simultaneously capture both overapproximations
(may-style) and underapproximations (must-style) of program behaviors,
allowing safety and reachability specifications to be compositionally
combined within a single type discipline. This model gives rise to a
type system in which type qualifiers are equipped with a modality
indicating whether they should be interpreted as an over- or
under-approximate specification.  The type system naturally supports
both deterministic and nondeterministic higher-order programs, captures
both may- and must-style function summaries~\cite{GN+10}, and can be
applied to a range of verification tasks, including
incorrectness-style reasoning~\cite{OHP19} and proving completeness of
test-input generators~\cite{CoverageType}.

To illustrate, consider the implementation of Peano division from
Rocq's standard library:
\begin{align}
  \lambda x.\lambda y. \zif{y \leq 0}{x}{\Code{fst~(divmod}~x~(y - 1)~0~(y - 1))}
  \tag{\Code{div}}\label{ex:div}
\end{align}
A client of \ref{ex:div} may want to know that (a) on positive
inputs, its result is never greater than its first argument (a safety
property); and (b) for every positive number $n$, there exist positive
inputs $x$ and $y$ such that \ref{ex:div} $x~ y = n$ (a reachability
property). Conventional refinement type systems, which interpret
refinements as overapproximations, can verify the former, but not the
latter: e.g., a function that always returns $1$ satisfies
(a), but not (b). Conversely, an underapproximate interpretation can establish (b),
but would also admit an implementation that always returns $y$, potentially violating (a). Our
type system, in contrast, captures both kinds of properties as types:\footnote{As we show later,
  these two types can be combined via intersection to simultaneously verify both
  requirements.}
\begin{align}
  & x{:}\ort{\Nat}{\vnu > 0} \sarr y{:}\ort{\Nat}{\vnu > 0 } \sarr \ort{\Nat}{x \geq \vnu} \tag{$\tau_a$}\label{ex:tauA}\\
  & x{:}\urt{\Nat}{\vnu > 0} \sarr y{:}\urt{\Nat}{\vnu > 0} \sarr \urt{\Nat}{\vnu > 0 } \tag{$\tau_b$}\label{ex:tauB}
\end{align}\noindent
The result type of \ref{ex:tauA} uses $\ort{b}{\phi}$ to express that
the qualifier $\phi$ should be interpreted as a safety constraint: a
term with this type cannot produce any value outside the set of values
the type denotes.  Conversely, \ref{ex:tauB} uses $\urt{b}{\phi}$ to
capture a reachability constraint: a term with this type must produce
every value found in the set the type denotes.

The intuitive meaning of each modality changes subtly when a
refinement appears in a typing context instead of a result type. To
establish the typing judgment $x{:} \ort{b}{\phi} \vdash e : \tau$,
for example, our type system treats any free occurrences of $x$
\emph{demonically}, since the surrounding environment is free to
choose some value of $x$ consistent with $\phi$ that makes $e$ violate
the constraints defined by $\tau$. Conversely, to establish
$x{:}\urt{b}{\phi} \vdash e : \tau$, $x$ is treated \emph{angelically}:
the surrounding environment is guaranteed to assign $x$ every value
consistent with $\phi$, allowing the type system to assume any such
value is available when verifying $e$ against $\tau$.

The availability of first-class functions adds additional
complications. Higher-order functions like
$x{:} \tau_x \vdash \lambda y. e : y{:}\tau_y \sarr \tau$, for
example, naturally expose two distinct contexts: the first,
$x{:} \tau_x$, captures the type of the free variable $x$, while the
second, $y{:} \tau_y$, provides assumptions about the function's
arguments at the point it is applied. Our system must manage these
contexts separately, because the instantiations of the free variables
and parameters of a function can occur at different points in a
program's evaluation. To illustrate, consider the following (failing)
type judgement:
\begin{align*}
    x{:}\urt{\Nat}{\vnu > 0} &\vdash  \lambda f. f\;x ~~:~~ f{:}(\urt{\Nat}{\vnu > 0}\sarr \urt{\Nat}{\vnu > 0}) \sarr \urt{\Nat}{\vnu > 0} \quad\judgfail
\end{align*}\noindent
The higher-order function being typed here takes a function $f$ and
directly applies it to $x$. The type binding for $x$ in the type
context assumes it can reach every positive number.  Without
constraining the type of $f$ to account for the free variable $x$ in
the expression's closure, however, we cannot guarantee that an
arbitrary function supplied for $f$ will guarantee the reachability
requirement of this expression's result.  For example, applying this
expression to the function $\lambda y.x - y$ yields the constant
function that always returns $0$.
\begin{align*}
  & \zlet{x}{e_x}{\boxed{(\lambda f. f\;x)\;(\lambda y.x - y)}} \;\hookrightarrow^*
    \; \zlet{x}{e_x}{\boxed{x - x}} \;\hookrightarrow^*\; 0
\end{align*}\noindent
In contrast, applying $\lambda f. f\;x$ to the identity function,
$\lambda y.y$, yields the term $x$, which can be made to reach any
particular positive number $n$ by choosing $x$ to be $n$.  The problem
is that the ``captured context'' of the function $\lambda y.x - y$
contains $x$ \emph{and} that this function is supplied $x$ as its
argument. Thus, in order to assign $\lambda f. f\;x$ the desired
result type, the type of $f$ must enforce that its parameter $y$ can
be chosen independently of $x$. This requires our type system to
reason explicitly about dependencies between a function's parameters
and its captured variables when both are treated angelically.

The previous example showed how dependencies between captured
variables and function parameters can break compositional reachability
reasoning.  The example below illustrates how a similar problem can
arise dynamically, during program execution. When the result of an
intermediate function application depends on a captured variable, it
can constrain the angelic choices available to subsequent applications
in a way that also breaks compositional reasoning:
\begin{align*}
  y{:}\urt{\Nat}{\vnu > 0} \vdash \lambda f. f(f\;y) : f{:}(\urt{\Nat}{\vnu > 0}\sarr \urt{\Nat}{\vnu > 0}) \sarr \urt{\Nat}{\vnu > 0} \quad \judgfail
\end{align*}\noindent
A counter-example to the validity of this judgement is given below:
{ \setlength{\jot}{-1pt} %
\begin{align*}
  & \zlet{x}{e_x}{\zlet{y}{e_y}{\boxed{(\lambda f. f\;(f\;y))\;g}}} \;\hookrightarrow^*\;
  \quad  \text{\hfill\textcolor{gray}{(where $g \doteq \lambda y.\zif{y = 1}{x}{2}$)}}
  \\& \zlet{x}{e_x}{\boxed{g\;(g\;1)}} \;\hookrightarrow^*\;  \hspace{1.38in}
  \text{\textcolor{gray}{\hfill(we assume angelically that $e_y \hookrightarrow^* 1$)}}
    \\& \zlet{x}{e_x}{\boxed{g\;x}} \;\hookrightarrow^*\; \zlet{x}{e_x}{\boxed{\zif{x = 1}{x}{2}}} \;\hookrightarrow^*\; \boxed{1 \text{ or } 2}
\end{align*}
}
\noindent
Angelically choosing $y = 1$, the first application ($g\; 1$) returns
$x$. This result is then passed to $g$ again, but since $x$ is a
captured variable rather than a fresh angelic choice, the second
application $(g\; x)$ can only return 1 or 2, and so the nested
application fails to reach any positive number greater than 2. Thus,
although the individual applications of $g$ in the original program
appear to satisfy its reachability specification, their composition
does not, again invalidating compositional reasoning.

To solve the context dependency issues raised above, we equip typing
contexts with \emph{capabilities} that are treated as \emph{resources}
that mediate demonic or angelic choices as necessary.  Concretely, a
capability associated with a typing context $\Gamma$ is a set of
environments consistent with $\Gamma$, representing the range of
variable bindings available for demonic or angelic choices.  We then
introduce a \emph{resource algebra} to compose and destruct these
capabilities in a principled way.  Specifically, the algebra ensures
that capabilities can only be combined and decomposed in a way that
faithfully reflects the dependency structure between variable bindings
in the typing context, as illustrated in the examples given above.
This approach enables us to handle context independence analogously to
the way bunched implications (BI)~\cite{OHP99} handle resource
separation, lifting the additive/multiplicative conjunction
distinction from the level of propositions to the level of type
contexts.

This framing naturally leads to the notion of a \emph{context type}
that unifies safety and reachability reasoning in a functional
language.  Specifically, we allow type contexts to be related via
dependent ($\Gamma_1 \ctcomma \Gamma_2$) and independent
($\Gamma_1 \ctstar \Gamma_2$) conjunction.  The type constructor
$\ctsarr$ internalizes dependent conjunction, typing functions whose
argument context is entangled with its closure context, while
$\ctwand$ internalizes independent conjunction, typing functions whose
argument context does not depend on the closure, directly reflecting
whether angelic choices for the argument are constrained by the
closure.

This bunched type system enables the tracking of complex data-flow
dependencies and informs the angelic choices that are allowed in a
typing judgement.  Specifically, since contexts can be combined via
context combinators, the structure of a context is naturally a binary
tree whose internal nodes record how sub-contexts are combined,
enabling frame-based local reasoning.  For example, to establish that
\ref{ex:div} can produce all positive numbers, we can type it against
the following type signature that provides a more precise version of $\tau_b$
using $\ctwand$ to make explicit that $x$ and $y$ must be chosen independently:
\begin{align*}
  & \vdash \ref{ex:div} ~:~ x{:}\urt{\Nat}{\vnu > 0} \ctsarr y{:}\urt{\Nat}{\vnu > 0} \ctwand \urt{\Nat}{\vnu > 0} \quad \judgok
\end{align*}\noindent
Here, the type constructor ($\ctwand$) requires the arguments for $x$
and $y$ to be independent, guaranteeing that each angelic choice can
be made regardless of the other.

\paragraph{Contributions.}
The paper makes the following contributions:\footnote{The type system,
  logic, and metatheory have been fully mechanized in Rocq, and are
  included in the supplemental material.}
\begin{itemize}[leftmargin=*]
\item We interpret the capability of a context to make angelic or
  demonic choices as a resource, and introduce a resource algebra to
  express compositional operations over this capability.
\item Building on this algebra, we define an associated modal
  \emph{context logic} that supports entangled and independent mixing
  of safety/demonic and reachability/angelic modalities.
\item We introduce \emph{context types}, a bunched modal type system
  interpreted in our context logic, to support both safety and
  reachability reasoning for deterministic functional programs.
\item We further generalize context types to a functional language
  extended with nondeterminism, where the result context is allowed to
  be disjoint from the input context.
\end{itemize}

The remainder of the paper is organized as follows.  We begin with an
example-driven tour of our type system.
Section~\ref{sec:context-typing} formalizes this informal overview in
terms of a typed $\lambda$-calculus core language.
Section~\ref{sec:meta} defines a denotational semantics and metatheory
for this type system via a bunched implication resource capability
algebra.  Section~\ref{sec:extension} extends the language with
nondeterminism, and describes supporting modifications to the logic
and denotational semantics.  Section~\ref{sec:case} presents several
case studies for which context types are particularly well-aligned.  A
brief discussion on potential ways to realize effective verification
tools based on this framework is given in
Section~\ref{sec:discussion}.  Related work and conclusions follow in
Sections~\ref{sec:related} and~\ref{sec:conclusion}.

\section{Overview}
\label{sec:overview}
Before presenting the full details of our type system, we begin with a
tour of its key features and how they enable both safety and
reachability reasoning.

\paragraph{Demonic and Angelic Modalities}

As in other refinement type systems, a context type refines a base
type $b$ with a qualifier $\phi$, a sentence in first-order
logic. Every context type is equipped with a modality — either demonic
($\ort{b}{\phi}$) or angelic ($\urt{b}{\phi}$) — indicating how its
qualifier should be interpreted.  Thus, a context type describes the
safety and reachability properties of an expression. As discussed in
Section~\ref{sec:intro}, when ascribed to an expression $e$, the demonic
context type $\ort{b}{\phi}$ allows $e$ to (\emph{at most}) acquire
any value $v$ that satisfies its qualifier $\phi$.  Conversely, an
angelic context type $\urt{b}{\phi}$ stipulates that $e$ must
(\emph{at least}) acquire every value satisfying $\phi$.

To illustrate this difference, consider the following typing
judgements which ascribe different context types to the constant term
$1$:
\begin{align*}
  &\vdash 1 : \ort{\Nat}{\top}\;\judgok
  && \vdash 1 : \urt{\Nat}{\top}\;\judgfail
  \\&\vdash 1 : \ort{\Nat}{\vnu > 0}\;\judgok
  && \vdash 1 : \urt{\Nat}{\vnu > 0}\;\judgfail
  \\&\vdash 1 : \ort{\Nat}{\vnu = 1}\;\judgok
  && \vdash 1 : \urt{\Nat}{\vnu = 1}\;\judgok
\end{align*}\noindent
The judgements on the left assert \emph{safety} properties, similar to
a standard refinement type system, where the constant $1$ needs to
satisfy the corresponding qualifier. The judgements on the right, in
contrast, describe \emph{reachability} constraints: since the constant
$1$ cannot reach every natural number or positive number, the first
two judgements are invalid. The last (singleton) type precisely
captures the reachability properties the term exhibits. Under an empty
typing context, a deterministic term %
can only be assigned either an empty or singleton angelic type, as it
can evaluate to at most one value.  Even for deterministic terms,
however, richer reachability properties naturally arise when reasoning
within a broader program context. To do so, our type system
\emph{holistically} considers the \emph{set} of values that a term can
evaluate to, i.e., its image, under a restricted \emph{set} of
environments, or \emph{capability}, that is constrained by a typing
context.

Intuitively, a demonic refinement acts as an \emph{upper} bound: when
used as a result type, $\ort{b}{\phi}$ requires every value in the
image of that term to satisfy $\phi$; in a type context, it only
admits capabilities in which every environment maps the variable to a
value satisfying $\phi$. An angelic refinement, in contrast, acts as a
\emph{lower} bound: as a result type, $\urt{b}{\phi}$ requires the
image of a term to contain every value satisfying $\phi$; in a type
context, it forces capabilities to at least include environments
mapping the variable to each value that satisfies $\phi$. To
illustrate, %
consider the following typing judgements involving non-singleton
reachability properties: %
\begin{align*}
  x{:}\urt{\Nat}{\vnu > 3} \vdash x - 3 : \urt{\Nat}{\vnu >
  0}\;\judgok \quad & \quad
  x{:}\urt{\Nat}{\vnu > 4} \vdash x - 3 : \urt{\Nat}{\vnu > 0}\;\judgfail
\end{align*} %
\noindent
Clearly, for any particular positive number $n$, the term $x - 3$ will
evaluate to that $n$, assuming $x$ can be assigned $n + 3$.  The
angelic types of $x$ in each typing context place different
constraints on the set of possible assignments to $x$ in a capability:
under the first, for every $n$ greater than $3$, the capability must
contain an environment in which $x$ is $n$; the second allows a
capability which does not include a mapping of $x$ to $4$, and thus
cannot guarantee the image of $x - 3$ includes $1$.

Switching the modality on the type of $x$ inverts the direction of
these bounds, thereby restricting the set of assignments to $x$ that
are impossible. As a consequence, the reachability guarantees that our
type system can provide about deterministic terms involving only
demonically qualified variables are somewhat limited:
\begin{align*}
  x{:}\ort{\Nat}{\vnu > 2} \vdash x - 3 : \urt{\Nat}{\vnu > 0}\;\judgfail %
\end{align*} %
Observe there are many singleton capabilities containing an
environment in which $x$ is assigned a value greater than $2$, but
$x - 3$ can only reach a single result value under such a capability!
Similarly, our type system can provide somewhat limited safety
guarantees when only equipped with reachability assumptions:
\begin{align*}
  x{:}\urt{\Nat}{\vnu > 4} \vdash x - 3 : \ort{\Nat}{\vnu > 0}\;\judgfail
\end{align*} %
The image of $x - 3$ includes a non-positive value under any
capability containing an environment where $x$ is $3$, and the lower
bound provided by the angelic type of $x$ in the typing judgement
above does not preclude such an environment.

\paragraph{Subtyping}

The subtyping relationship between two angelic or two demonic
refinements is similar to other contemporaneous refinement type
systems~\cite{LiquidHaskell, CoverageType}, where the subtype
relationship between demonic refinements is covariant with the
qualifier, and that of angelic refinements is contravariant with the
qualifier. These relationships align with the interpretation of
angelic and demonic modalities as lower and upper bounds: a term
which can reach all positive numbers can also reach all numbers
greater than 1, while a term that always safely evaluates to a number
greater than 1 is also guaranteed to always evaluate to a positive
number.
\begin{align*}
  &\ort{\Nat}{\vnu > 1} \subty \ort{\Nat}{\vnu > 0}\;\judgok
  &&\urt{\Nat}{\vnu > 1} \subty \urt{\Nat}{\vnu > 0} \;\judgfail \\
  & \ort{\Nat}{\vnu > 0} \subty \ort{\Nat}{\vnu > 1}  \;\judgfail
  && \urt{\Nat}{\vnu > 0} \subty \urt{\Nat}{\vnu > 1}\;\judgok
\end{align*} %

The subtyping relationship between angelic and demonic refinement
types is more nuanced. No subtyping relationship exists between two
arbitrary refinements with different modalities: %
\begin{align*}
  &\ort{\Nat}{\vnu > 0} \subty \urt{\Nat}{\vnu > 0}\;\judgfail
  & &\urt{\Nat}{\vnu > 0} \subty \ort{\Nat}{\vnu > 0} \;\judgfail
\end{align*} %
\noindent
Furthermore, it is only possible to switch the modality of a
refinement when its qualifier is a singleton, or when the resulting
refinement type trivially holds, i.e., $\ort{b}{\top}$ or
$\urt{b}{\bot}$: %
\begin{align*}
  &\ort{\Nat}{\vnu = 3} \subty \urt{\Nat}{\vnu =
    3}\;\judgok
  & & \urt{\Nat}{\vnu = 3} \subty \ort{\Nat}{\vnu =
      3}\;\judgfail \\
  & \urt{\Nat}{\vnu > 0} \subty \ort{\Nat}{\top}\;\judgok
  & &\ort{\Nat}{\vnu > 0} \subty \urt{\Nat}{\bot}\;\judgok
\end{align*}\noindent
When the qualifier is a singleton, safety entails reachability ---
there is only one possible value, so the demonic constraint (the value
may only be 3) and the angelic constraint (the value 3 must be
reachable) make identical demands on the term. Note that the converse
direction does not hold, however. Converting to a supertype with a
trivial qualifier, in contrast, simply throws away the bounds on a
term entirely.

\paragraph{Type Context Composition}
When they appear in a typing context, the bounds imposed by our two
modalities afford us different reasoning capabilities: when
establishing the reachability properties of a term, for example, an
angelic refinement lets us \emph{choose} a favorable assignment for a
variable in that term, while a demonic refinement allows us to
\emph{rule out} certain assignments to that variable when showing that
a term is safe.
Importantly, the bounds imposed by the bindings of individual
variables in a typing context can interact with each other, altering
the capabilities afforded by the typing context as a whole. To
illustrate, consider the following pair of program contexts:
\begin{align}
  &\zlet{x}{\Code{gen_{nat} () }}{\zlet{y}{x + 1}{\square}}
    \tag{C$_\mathit{en}$}\label{ex:EnCon} \\
  &\zlet{x}{\Code{gen_{nat} () }}{\zlet{y}{\Code{gen_{nat}()}}{\square}}
    \tag{C$_\mathit{dis}$}\label{ex:DisjCon}
\end{align}\noindent
Both contexts use $\Code{gen_{nat} ()}$ to (nondeterministically)
generate an arbitrary value for $x$. The first context then sets $y$
based on the value of $x$, while the second opts to generate a fresh
value for $y$ instead. The values bound to $x$ and $y$ in either
context can all reach every possible positive number, as can the
program \ref{ex:DisjCon}[$x - y$].  This is not true of
\ref{ex:EnCon}[$x - y$], however, because the values of $x$ and $y$ in
this program are \emph{entangled} since the values $y$ can yield are
tied to $x$'s.
\begin{align*}
  &\emptyset \vdash \text{\ref{ex:DisjCon}}[x] : \urt{\Nat}{\vnu > 0}
    \;\judgok \quad
  &\emptyset \vdash \text{\ref{ex:EnCon}}[x] : \urt{\Nat}{\vnu > 0}
    \;\judgok  \\
  & \emptyset \vdash \text{\ref{ex:DisjCon}}[y] : \urt{\Nat}{\vnu > 0}
  \;\judgok \quad
  &\emptyset \vdash \text{\ref{ex:EnCon}}[y] : \urt{\Nat}{\vnu > 0}
    \;\judgok \\
  & \emptyset \vdash \text{\ref{ex:DisjCon}}[x - y] : \urt{\Nat}{\vnu
    > 0} \;\judgok \quad & \emptyset \vdash \text{\ref{ex:EnCon}}[x - y] : \urt{\Nat}{\vnu > 0} \;\judgfail
\end{align*}\noindent
To deal with this situation, our type system includes two constructors
over typing contexts, ``$\ctcomma$'' and ``$\ctstar$'', each of which
affords different capabilities when reasoning about, e.g., subtyping
relationships. Informally, $\Gamma_1 \ctstar \Gamma_2$ means that the
capabilities of $\Gamma_1$ and $\Gamma_2$ are independent; when either
context includes an angelic type binding, we are free to ``choose'' a
favorable assignment for the corresponding variable independent of the
other context.  Dually, $\Gamma_1 \ctcomma \Gamma_2$ means that the
capabilities of $\Gamma_1$ and $\Gamma_2$ are dependent; the angelic
choices afforded by $\Gamma_2$ may be constrained by choices for
variables in $\Gamma_1$.  Consider the following two subtyping
judgments reflecting the typing of $\square $ in $C_{\mathit{dis}}$
and $C_{\mathit{en}}$, resp:
\begin{align*}
  &x{:}\urt{\Nat}{\vnu \geq 0} \ctstar y{:}\urt{\Nat}{\vnu \geq 1}
    \vdash \urt{\Nat}{\vnu = x - y} \subty \urt{\Nat}{\vnu > 0}
    \;\judgok \tag{$\Gamma_\mathit{dis}$}\label{eqn:Ctxdis} \\
  &x{:}\urt{\Nat}{\vnu \geq 0} \ctcomma y{:}\urt{\Nat}{\vnu \geq 1}
    \vdash \urt{\Nat}{\vnu = x - y} \subty \urt{\Nat}{\vnu > 0} \;\judgfail
    \tag{$\Gamma_\mathit{en}$}\label{eqn:Ctxen}
\end{align*}\noindent
Under \ref{eqn:Ctxdis}, we can freely choose both $2$ for $y$
\emph{and} an arbitrary value larger than $2$ for $x$, from which the
subtyping relationship follows. Under \ref{eqn:Ctxen}, however, the
possible values we can choose for $x$ and $y$ are constrained by the
\emph{entangled} conjunction, ``$\ctcomma$'': once a value is chosen
for $x$, $y$ can no longer be freely chosen. Concretely, the angelic
guarantee for $y$ under ``$\ctcomma$'' only promises that $y$ can
reach every value satisfying $\vnu > 1$ \emph{for some fixed
  instantiation of $x$}.  Since the subtyping goal requires producing
every positive number as the difference of $x$ and $y$, $x - y$, and
this requires pairing each candidate value of $x$ with an
independently chosen $y$, a dependent context cannot be used to supply
such pairs.  This is evidenced by $C_{en}$, which always collapses its
result to 0, rather than all positive numbers.  The ``\ctcomma'' in
the context $\Gamma \ctcomma x{:}\tau_x \ctcomma y{:}\tau_y$ works
like additive logic conjunction in BI, and both
$\Gamma\ctcomma x{:}\tau_x \ctcomma y{:}\tau_y \vdash x{:}\tau_x$
and
$\Gamma\ctcomma x{:}\tau_x \ctcomma y{:}\tau_y \vdash y{:}\tau_y$
are valid. Importantly, however, it does not guarantee that the
modalities of $x$ and $y$ hold \emph{at the same
  time}. \footnote{Unlike a linear type system, which would
  \emph{consume} the capability to angelically assign $x$ when
  producing the capability to angelically assign $y$, our system
  instead \emph{entangles} these capabilities, allowing the body
  expression $e$ to use both $x$ and $y$ while still tracking their
  dependency. This formulation aligns with the intuition of bunched
  typing~\cite{BunchedType}. %
}  For example,
\begin{align*}
  &x{:}\urt{\Nat}{\vnu > 0} \ctstar (y{:}\urt{\Nat}{\vnu > 0} \ctcomma z{:}\urt{\Nat}{\vnu > 0}) \vdash x - y : \urt{\Nat}{\vnu > 0} \;\judgok
  \\&x{:}\urt{\Nat}{\vnu > 0} \ctstar (y{:}\urt{\Nat}{\vnu > 0}
  \ctcomma z{:}\urt{\Nat}{\vnu > 0}) \vdash x - (z + y) : \urt{\Nat}{\vnu > 0} \;\judgok
  \\&x{:}\urt{\Nat}{\vnu > 0} \ctstar (y{:}\urt{\Nat}{\vnu > 0} \ctcomma z{:}\urt{\Nat}{\vnu > 0}) \vdash y - z : \urt{\Nat}{\vnu > 0} \;\judgfail
\end{align*}\noindent
In these judgements, the context tracks variable dependency,
specifying that $x$ is independent from $y$ and $z$, while $y$ and $z$
are entangled with each other.  Specifying independence in this way is
sufficient to verify reachability for $x - y$ and $x - (z + y)$, since
$x$ can be angelically chosen independently of both $y$ and $z$.
However, type-checking $y - z$ against the type $\urt{\Nat}{\vnu > 0}$
fails because $y$ and $z$ are entangled: no single angelic choice can
simultaneously make their difference reach every positive number.

Notably, angelic and demonic refinements can also be entangled in a
way that impacts the capabilities afforded by a type context, as the
following pair of judgments illustrate:
\begin{align*}
  &x{:}\urt{\Nat}{\vnu \geq 0} \ctcomma y{:}\ort{\Nat}{\vnu \geq 1} \vdash x - y : \urt{\Nat}{\I{even}(\vnu)} \;\judgfail
  \\&x{:}\urt{\Nat}{\vnu \geq 0} \ctstar y{:}\ort{\Nat}{\vnu \geq 1} \vdash x - y : \urt{\Nat}{\I{even}(\vnu)} \;\judgok
\end{align*}\noindent
The first typing context stipulates that the value of $y$ is positive,
but that value is allowed to depend on the value $x$ --- the hole in
\ref{ex:EnCon} can also be typed under this typing
context. Accordingly, the expression \ref{ex:EnCon}$[x - y]$ cannot
reach every even number. Contrast this with a variant of this context
that instead assigns \emph{some} constant value $n$ to $y$:
\begin{align}
  &\zlet{x}{\Code{gen_{nat} () }}{\zlet{y}{n}{\square}}
    \tag{C$_{n}$}\label{ex:DisjSafeCon}
\end{align}
Regardless of the particular value of $n$, for any given even number
$m$, choosing $x = m + n$ allows \ref{ex:DisjSafeCon}$[x - y]$ to
reach that number.

\paragraph{Function Types}
Entangled ($\ctsarr$) and independent
function types ($\ctwand$) are adjunctions of the corresponding type
context composition operators:
\begin{align*}
  \Gamma \vdash e_f : x{:}\tau_x \ctsarr \tau \quad &\text{ iff } \quad \Gamma \ctcomma x{:}\tau_x \vdash (e_f\;x) : \tau
  \\\Gamma \vdash e_f : x{:}\tau_x \ctwand \tau \quad &\text{ iff } \quad \Gamma \ctstar x{:}\tau_x \vdash (e_f\;x) : \tau
\end{align*}\noindent
Thus, ($\Code{-}$) can be given the following types:
\begin{align*}
  &\vdash (-) : x{:}\urt{\Nat}{\vnu > 0} \ctwand y{:}\urt{\Nat}{\vnu > 0} \ctwand \urt{\Nat}{\vnu > 0}
  \\&\vdash (-) : x{:}\urt{\Nat}{\vnu > 0} \ctsarr y{:}\urt{\Nat}{\vnu > 0} \ctwand \urt{\Nat}{\vnu > 0}
\end{align*}\noindent
Notably, however, it cannot have the following type:
\begin{align*}
  &\vdash (-) : x{:}\urt{\Nat}{\vnu > 0} \ctwand y{:}\urt{\Nat}{\vnu > 0} \ctsarr \urt{\Nat}{\vnu > 0}
\end{align*}\noindent
Doing so would require proving the following type judgement:
\begin{align*}
  &\emptyset \ctstar x{:}\urt{\Nat}{\vnu > 0} \ctcomma y{:}\urt{\Nat}{\vnu > 0} \vdash x - y : \urt{\Nat}{\vnu > 0} \;\judgfail
\end{align*}\noindent
As we saw earlier, a type context in which $x$ and $y$ are entangled
restricts the angelic choices that can be made, preventing us from
establishing that $x - y$ can produce every positive number.

Higher-order functions introduce additional complexity. Consider the
following example:
\begin{align*}
  &x{:}\urt{\Nat}{\vnu > 0} \ctstar y{:}\urt{\Nat}{\vnu > 0} \vdash \zzlam{f}{x - (f\; (f\;y))} :
  \\&\quad  \underbrace{f{:}(\underbrace{y{:}\urt{\Nat}{\vnu > 0} \ctsarr}_\texttt{$y$ can depend on context during application} \urt{\Nat}{\vnu > 0}) \ctwand}_\texttt{$f$ is independent from $x$ and $y$} \urt{\Nat}{\vnu > 0}
\end{align*}\noindent
\cbnewadding{The two arrows serve distinct purposes: the inner
$\ctsarr$ permits the argument of each application of $f$ to depend
on $f$'s closure, whereas the outer $\ctwand$ keeps $f$ independent
of the surrounding bindings for $x$ and $y$.}
To make $x - (f\;(f\;y))$ reach every positive number, $x$ and
$f\;(f\;y)$ must be independent. To ensure this, one might expect that
all function arrows in the result type would need to be $\ctwand$, but
this would be incorrect. If $f$ had the fully independent type
$y{:}\urt{\Nat}{\vnu > 0} \ctwand \urt{\Nat}{\vnu > 0}$, then each
application of $f$ would require its argument to be independent from
$f$'s closure. While the first application $f\;y$ satisfies this, the
second application $f\;(f\;y)$ fails because the intermediate result
$f\;y$ depends on $f$. While $\ctwand$ guarantees that $f$'s argument
is not entangled with its closure, it makes no guarantees that $f$'s
result is also independent. Thus, while $(f\; y)$ is independent of
$y$, it may still depend on $f$'s closure, making it an invalid
argument in the outer application when equipped with this signature.

If $f$ is instead given the entangled function type
$y{:}\urt{\Nat}{\vnu > 0} \ctsarr \urt{\Nat}{\vnu > 0}$, its argument
can now depend on $f$'s closure, accommodating the nested
application. The function itself is then typed with $\ctwand$,
ensuring $f$'s body is not entangled with either $x$ or $y$ in the
outer context. By adjunction, this yields:
\begin{align*}
  &x{:}\urt{\Nat}{\vnu > 0} \ctstar
  \\&\quad (y{:}\urt{\Nat}{\vnu > 0} \ctstar
  f{:}(y{:}\urt{\Nat}{\vnu > 0} \ctsarr \urt{\Nat}{\vnu > 0}))
  \vdash x - (f\; (f\;y)) : \urt{\Nat}{\vnu > 0}
\end{align*}\noindent
Since $x$ is independent from both $y$ and $f$, it is also independent
from $f\;(f\;y)$, thus guaranteeing that $x - (f\;(f\;y))$ can reach
every positive number.

These examples illustrate a flexibility that, to our knowledge, is
absent from prior refinement type systems: demonic and angelic
qualifiers may appear in arbitrary combination throughout a type, and
the entangled ($\ctsarr$) and independent ($\ctwand$) function type
constructors may likewise be chosen independently at each argument
position.

\paragraph{From Reachability to Safety}
As discussed above, the sort of safety that can be directly derived
from reachability assumptions in the typing context is quite limited
--- ascribing a demonic refinement to a term constrains \emph{all} of
its possible executions, while an angelic type binding only makes
assumptions about a \emph{subset} of the environments that a term may
be run in. It is possible, however, to derive meaningful safety
properties about the executions that an angelic refinement guarantees
exist. If we have assumed that $x$ must reach every even number, for
example, we know that the term $x \Code{+ 1}$ will always safely
evaluate to an odd number \emph{under those environments}. In order to
soundly establish upper bounds on program executions that are
guaranteed to exist, our type system includes a \emph{context sum}
type constructor $\ctoplus$ that explicitly identifies partitions of
the values a term must reach.

The typing judgement $\Gamma \vdash e : \tau_1 \ctoplus \tau_2$ states
that $e$ can be typed with $\tau_1$ and $\tau_2$ in some partitioning
of $\Gamma$, and that all capabilities consistent with $\Gamma$ are
accounted for by $\tau_1$ and $\tau_2$ --- no possibility is lost.
For example: \begin{align*}
  &x{:}\urt{\Nat}{\vnu < 10} \vdash x + 1 : \ort{\Nat}{odd(\vnu)} \ctoplus
    \ort{\Nat}{even(\vnu)} \;\judgok \notag \\
  &x{:}\urt{\Nat}{\top} \vdash \zif{x == 0}{3}{5} : \ort{\Nat}{\vnu <
    4}\ctoplus \ort{\Nat}{\vnu > 4} \;\judgok
\end{align*}\noindent
In both judgements, the result type uses $\ctoplus$ to explicitly
demarcate two subsets of the term's executions. The first example uses
$\ctoplus$ to establish that the angelic refinement of $x$ in the
typing context ensures that $x + 1$ must evaluate to at least one even
and at least one odd number, without stipulating what those numbers
are. In the second example, $\ctoplus$ is used to distinguish between
the two possible control flows through the program.

Importantly, while
$\ort{\Nat}{\vnu < 4} \ctoplus \ort{\Nat}{\vnu > 4}$ is a subtype of
$\ort{\Nat}{\vnu < 4 \lor \vnu > 4}$ (both types can only be ascribed
to a term that never evaluates to $4$), the reverse is not true: the
disjunctive refinement type loses the information that a term has two
distinct feasible control flow paths, one whose result is always
greater than $4$ and another whose result is always less than
$4$.\footnote{In this
  sense, the $\ctoplus$  operator plays a role analogous to the commutative
  monoid disjunction in Outcome Logic~\cite{ZDS23}.} This ability also
allows us to focus on the safety properties of one particular feasible
program path, assigning all remaining paths the $\top$ type: %
\begin{align*}
  &x{:}\urt{\Nat}{\top} \vdash \zif{x > 0}{5}{3} : \ort{\Nat}{\vnu >
    4} \ctoplus \ort{\Nat}{\top} \quad \judgok
    \tag{\Code{e}$_{\oplus\top}$}\label{ex:oplusWeak}
\end{align*}\noindent
The use of $\tau \ctoplus \ort{b}{\top}$ expresses an \emph{affine}
style of control flow analysis when deriving information from angelic
facts: like affine logic, which allows resources to be used at most
once (and hence discarded), this style allows the type system to
discard uninteresting branches by weakening them to $\ort{b}{\top}$,
imposing a precise safety constraint on the branches that a typing
context guarantees exist while placing no constraint on the others.

\cbnewadding{The $\ctoplus$ constructor can also be used to combine
  type contexts.  When used in a type judgment, it records a partition
  of the available environments, just as a sum in the result type
  records a partition of executions.  For example, this combinator can
  be used to decompose the angelic context $x{:}\urt{\Nat}{\top}$ into
  the contexts $x{:}\urt{\Nat}{\vnu = 0}$ and
  $x{:}\urt{\Nat}{\vnu > 0}$.  This partitioning aligns the typing
  context of \ref{ex:oplusWeak} with its result type: the latter
  context describes environments that reach the true branch, whose
  result has type $\ort{\Nat}{\vnu > 4}$, while the former describes
  those that reach the false branch, which has the type
  $\ort{\Nat}{\top}$.}

\begin{figure}[!t]
  {\small
    \begin{alignat*}{2}
      \text{\textbf{Variables }}& \quad &\quad& x, y, z, \vnu, ...
      \\\text{\textbf{Data Constructors }}& \quad &d ::= \quad & () ~|~ \Code{true} ~|~ \Code{false} ~|~ \Code{O} ~|~ \Code{S} ~|~ \Code{Cons} ~|~ \Code{Nil} ~|~ ...
      \\\text{\textbf{Operations}}& \quad & \primop ::= \quad & d ~|~ {+} ~|~ {-} ~|~ {==} ~|~ {<} ~|~ {\leq} ~|~ ...
      \\ \text{\textbf{Values}}& \quad  & v ::= \quad & d\ \overline{v} ~|~ x ~|~ \zlam{x}{s}{e} ~|~ \zfix{f}{s}{x}{s}{e}
      \\ \text{\textbf{Expressions}}& \quad & e ::=\quad & v ~|~ \zlet{x{:}s}{e}{e} ~|~ \primop\ \overline{v} ~|~  v\;v ~|~ \match{v} \overline{d\ \overline{y} \to e}
      \\\text{\textbf{Base Types}}& \quad & b  ::= \quad &  \Code{unit} ~|~ \Code{bool} ~|~ \Code{nat} ~|~ \List[b] ~|~ ...
      \\\text{\textbf{Basic Types}}& \quad & s  ::= \quad &  b ~|~ s\sarr s
      \\\text{\textbf{Basic Type Context}}& \quad & \Sigma ::= \quad &  \emptyset ~|~ x{:}s ~|~ \Sigma, \Sigma
    \end{alignat*}
  }
  \vspace{-18pt}
  \caption{Syntax of \langname{}}
  \label{fig:core-lang-syntax}
  \vspace{-6pt}
  \end{figure}

\section{Context Type System}
\label{sec:context-typing}

We formalize our proposed context type system for the core language
(\langname{}) shown in \autoref{fig:core-lang-syntax}.  This language
is a standard call-by-value lambda calculus with inductive datatypes,
pattern matching (we use conditional expressions as syntactic sugar
for \Code{match}), and recursive functions. Expressions are written in
monadic normal form (MNF)~\cite{JO94}, a variant of A-Normal Form
(ANF)~\cite{FSDF93} that allows nested let-bindings. All terms in
\langname{} are deterministic, a restriction that we lift in
\autoref{sec:extension}.  The language's basic types (i.e., $s$)
consist of base types ($b$) and function types (i.e., $s\sarr
s$). Both \textbf{let}-expressions and function abstractions
explicitly annotate bound variables with a basic type (e.g.,
$\zlam{x}{s}{e}$); we omit these annotations for brevity, when
appropriate.  A basic type context $\Sigma$ maps variables to their
basic types; we assume that all bound variables in $\Sigma$ are
distinct.

\begin{figure}[h]
  { \small
  \vspace{-10pt}
    \begin{alignat*}{2}
      \text{\textbf{Qualifiers}}& \quad & \phi ::= \quad & v ~|~ x ~|~ \primop\;\overline{v} ~|~ \bot ~|~ \top  ~|~ \neg \phi ~|~ \phi \land \phi ~|~ \phi \lor \phi ~|~  \phi \impl \phi ~|~ \forall x{:}b. \phi ~|~ \exists x{:}b. \phi
      \\ \text{\textbf{Refinement Types}}& \quad & \tau ::= \quad &
      \ort{b}{\phi} ~|~ \urt{b}{\phi} ~|~ \tau \ctinter \tau ~|~ \tau
      \ctoplus \tau ~|~ x{:}\tau \ctsarr \tau ~|~ x{:}\tau \ctwand
      \tau %
      \\ \text{\textbf{Type Contexts}}& \quad & \Gamma ::= \quad & \emptyset ~|~ x{:}\tau ~|~ \Gamma \ctstar \Gamma ~|~ \Gamma \ctcomma \Gamma ~|~ \Gamma \ctoplus \Gamma
    \end{alignat*}}
  \vspace{-18pt}
  \caption{Context Type Syntax}
  \label{fig:context-types-syntax}
  \vspace{-6pt}
\end{figure}

The syntax of context types is shown in
\autoref{fig:context-types-syntax}. Like other refinement type
systems, \emph{context types} refine basic types with qualifiers
$\phi$; but each refinement is additionally given either a demonic or
angelic modality ($\otparenth{...}$ and $\utparenth{...}$,
resp.). Context types also include independent function types
($\ctwand$), sums ($\ctoplus$), and standard type intersections
($\ctinter$). While our type system shares similar intuitions as
bunched and refinement type systems, its technical development differs
significantly from both in order to support unified reasoning about
safety and reachability properties.

\begin{figure}[h!]
  {\small
  \vspace{-10pt}
  {\normalsize \begin{flalign*}
    &\text{\textbf{Well-formedness rules}} && \fbox{$\Sigma \wfvdash \tau \quad \Sigma \wfvdash \Gamma$}
  \end{flalign*}}
\mprooftr{23mm}
{$\Sigma \wfvdash \tau_x$ \quad $x \notin \Dom{\Sigma}$}
{WfSingle}
{$\Sigma \wfvdash x{:}\tau_x$}
\quad
\mprooftr{25mm}
{$\Sigma \wfvdash \Gamma_1$ \quad $\Sigma, \eraserf{\Gamma_1} \wfvdash \Gamma_2$}
{WfComma}
{$\Sigma \wfvdash \Gamma_1 \ctcomma \Gamma_2$}
\quad
\mprooftr{25mm}
{$\Sigma \wfvdash \Gamma_1$ \quad $\Sigma \wfvdash \Gamma_2$}
{WfStar}
{$\Sigma \wfvdash \Gamma_1 \ctstar \Gamma_2$}
}
\caption{Selected rules for well-formedness relations.}
\label{fig:auxiliary-rules}
\vspace*{-.1in}
\end{figure}

Our type system depends on a pair of well-formedness relations for
context types ($\Sigma \wfvdash \tau$) and type contexts
($\Sigma \wfvdash \Gamma$); select rules are shown in
\autoref{fig:auxiliary-rules}.\footnote{The full set of definitions
  are provided in the supplemental material.}  These standard
relations require, e.g., a context type to erase ($\eraserf{.}$) to a
basic type, type qualifiers to be well-typed in the basic type system,
and that any variables that can appear in qualifiers are unique
(\textsc{WfSingle}).  Importantly, in a well-formed entangled pair of
contexts, the second context is allowed to refer to variables in the
first (\textsc{WfComma}), while well-formed independent pairs of
contexts cannot refer to each other's variable bindings
(\textsc{WfStar}). In order for the type context
$\Gamma_0 \ctcomma (\Gamma_1 \ctstar \Gamma_2)$ to be well-formed, for
example, $\Gamma_0$ must be closed, and $\Gamma_1$ and $\Gamma_2$
cannot refer to variables in each other.

\begin{figure}[h!]
  {\small
  {\normalsize \begin{flalign*}
    &\text{\textbf{Subtyping rules}} && \fbox{$\Gamma \vdash \tau \subty \tau$ \quad $\Gamma \ctsubseteq{X} \Gamma$ }\quad
    &\text{\textbf{Typing rules}} && \fbox{$\Gamma \vdash e : \tau$}
  \end{flalign*}}
  \mprooftr{40mm}
  {$\Gamma \vdash e_1 : \tau_1 $ \quad $\Gamma \ctcomma x{:}\tau_1 \vdash e_2 : \tau_2$}
  {T-Let}
  {$\Gamma \vdash \zlet{x}{e_1}{e_2} : \tau_2$}
  \quad
  \mprooftr{40mm}
  {$\Gamma_1 \vdash e_1 : \tau_1 $ \quad $\Gamma_2 \ctstar x{:}\tau_1 \vdash e_2 : \tau_2$}
  {T-LetI}
  {$\Gamma_1 \ctstar \Gamma_2 \vdash \zlet{x}{e_1}{e_2} : \tau_2$}
  \\[.8em]
  \mprooftr{42mm}
  {$\eraserf{\tau_x} = s_x$ \quad $\Gamma \ctcomma x{:}\tau_x \vdash e : \tau$}
  {T-Lam}
  {$\Gamma \vdash \zlam{x}{s_x}{e} : x{:}\tau_x \ctsarr \tau$}
  \quad
  \mprooftr{42mm}
  {$\eraserf{\tau_x} = s_x$ \quad  $\Gamma \ctstar x{:}\tau_x \vdash e : \tau$}
  {T-LamI}
  {$\Gamma \vdash \zlam{x}{s_x}{e} : x{:}\tau_x \ctwand \tau$}
  \\[.8em]
  \mprooftr{40mm}
  {$\Gamma \vdash v_1 : x{:}\tau_x \ctsarr \tau$ \quad
  $\Gamma \vdash v_2 : \tau_x$ }
  {T-AppFun}
  {$\Gamma \vdash v_1\;v_2 : \tau[x \mapsto v_2]$}
  \quad
  \mprooftr{46mm}
  {$\Gamma_1 \vdash v_1 : x{:}\tau_x \ctwand \tau$ \quad
  $\Gamma_2 \vdash v_2 : \tau_x$  }
  {T-AppFunI}
  {$\Gamma_1 \ctstar \Gamma_2 \vdash v_1\;v_2 : \tau[x \mapsto v_2]$}
  \\[.8em]
  \mprooftr{30mm}
  {$\Gamma \vdash e : \tau_1$ \quad $\Gamma \vdash \tau_1 \subty \tau_2$}
  {T-Sub}
  {$\Gamma \vdash e : \tau_2$}
  \quad
  \mprooftr{40mm}
  {$\Gamma_2 \vdash e : \tau$ \quad $\Gamma_1 \ctsubseteq{\Code{FV}(e) \cup \Code{FV}(\tau)} \Gamma_2$}
  {T-CtxSub}
  {$\Gamma_1 \vdash e : \tau$}
  \quad
  \mprooftr{14mm}
  {$ $}
  {T-Var}
  {$x{:}\tau \vdash x : \tau$}
  \\[.8em]
  \mprooftr{34mm}
  {$\Code{Ty}(\primop) = \overline{x_i{:}\tau_i}\ctsarr \tau$ \quad
    $\forall i. \Gamma \vdash v_i : \tau_i$ }
  {T-AppOp}
  {$\Gamma \vdash \primop\;\overline{v_i} : \tau[\overline{x_i \mapsto v_i}]$}
  \quad
  \mprooftr{65mm}
  {$\Gamma \ctcomma x{:}\ort{b}{\phi} \ctcomma f{:}(x{:}\ort{b}{\phi \land \vnu \prec x} \ctsarr \tau) \vdash e : \tau$}
  {T-Fix}
  {$\Gamma \vdash \zzfix{f}{x}{e} : x{:}\ort{b}{\phi} \ctsarr \tau$}
  \\[.8em]
\cbnewadding{
  \begin{prooftree}
    \hypo{I \mathbin{\uplus} J = \{1,\ldots,n\} \quad I \neq \emptyset}
    \infer[no rule]1{\forall i \in I.\; \{\overline{y_i}\} \subseteq \Dom{\Gamma_i} \quad \text{and} \quad
      \Gamma_i \vdash
      d_i\;\overline{y_i} : \ort{b}{\vnu = v} \interty \urt{b}{\vnu = v}
    \quad \text{and} \quad
      \Gamma_i \vdash e_i : \tau_i
      }
    \infer[no rule]1{\forall j \in J.\;
      \left(\ctbigoplus_{i \in I} \Gamma_i\right)
      \ctcomma \overline{y_j{:}\ort{b_j}{\top}} \vdash
      d_j\;\overline{y_j} : \ort{b}{\vnu \neq v}}
    \infer1[\textsc{T-Match}]{\left(\ctbigoplus_{i \in I} \Gamma_i\right) \vdash
      \match{v} d_1\;\overline{y_1} \to e_1 \;\mid \cdots \; \mid \; d_n\;\overline{y_n} \to e_n
      : \ctbigoplus_{i \in I} \tau_i}
  \end{prooftree}
  }
}
  \caption{Declarative typing rules \langname{}}
  \label{fig:declarative-typing-rules}
  \vspace*{-.1in}
\end{figure}

The rules of our type system are shown in
\autoref{fig:declarative-typing-rules}. Every rule assumes well-formed
typing contexts and types; we elide these assumptions for clarity. The
typing rules for let-expressions, lambda functions, and applications
all have entangled and independent versions. In contrast to a standard
bunched type system, the entangled version of the $\mykw{let}$ rule
does not require an explicit additive conjunction in the type
context. Instead, \textsc{T-Let} directly entangles the new variable
binding with the existing context, without duplicating or removing
bindings.

\begin{example}[Typing Let-Expressions]
  To see why, consider the following let-expression:
\begin{align*}
    x{:}\ort{\Nat}{\vnu < 10} \vdash \zlet{y}{x - 3}{x - y} : \ort{\Nat}{\vnu \leq 3}
\end{align*}\noindent
Here, both the bound term $x - 3$ and the let-body $x - y$ refer to
the variable $x$. Applying the standard bunched typing rule for
let-expressions would require the following type context: %
\begin{align*}
  &\underbrace{x{:}\ort{\Nat}{\vnu < 10}}_{\texttt{for $x - 3$}} \ctcomma \underbrace{x{:}\ort{\Nat}{\vnu < 10}}_{\texttt{for $x - y$}} \vdash \zlet{y}{x - 3}{x - y} : \ort{\Nat}{\vnu \leq 3}
\end{align*} %
\noindent
Bunched type systems normally include a contraction rule that would
allow us to obtain such a context by duplicating the binding for $x$
in our target judgment. An entangled context with duplicate bindings
is not well-formed in our type system, so \textsc{T-Let} instead
directly combines this ``duplicate-then-consume'' pattern:
\cbnewadding{In this sense, \textsc{T-Let} simulates ordinary let
typing within our substructural type system: both the bound expression
and the body may use the existing context, without introducing
duplicate bindings into the context syntax.}
\begin{center}
  \begin{prooftree}
    \hypo{
      x{:}\ort{\Nat}{\vnu < 10} \vdash x - 3 : \ort{\Nat}{\vnu <
        7}}
    \infer[no rule]1{x{:}\ort{\Nat}{\vnu < 10} \ctcomma y{:}\ort{\Nat}{\vnu
        < 7} \vdash x - y : \ort{\Nat}{\vnu \leq 3}}
  \infer1[{\small \textsc{T-Let}}]{x{:}\ort{\Nat}{\vnu < 10} \vdash \zlet{y}{x - 3}{x - y} :
    \ort{\Nat}{\vnu \leq 3}}
  \end{prooftree}
\end{center}
\end{example}

Our system is equipped with subsumption rules for both types
(\textsc{T-Sub}) and type contexts (\textsc{T-CtxSub}). These rules
rely on \emph{semantic} relations which appeal to the denotations of
types and typing contexts; the definitions of both relations are given
in \autoref{sec:meta}. The intuition behind the subtyping relation
$\Gamma \vdash\tau_1 \subty \tau_2$ is standard, i.e., inhabitants of
the denotation of $\tau_1$ are also inhabitants of the denotation of
$\tau_2$ under the denotation of context $\Gamma$. The refinement
relation between typing contexts, $\Gamma_1 \ctsubseteq{X} \Gamma_2$
can be understood as ``any term that has type $\tau$ under $\Gamma_2$
can also be typed with $\tau$ under $\Gamma_1$''. The comparison is
confined to a particular set of variables $\textcolor{CBOrange}{X}$;
\textsc{T-CtxSub} restricts $\ctsubseteqA{}$ to the free variables of
a target term and type.  For example,
\begin{align*}
  x{:}\ort{\Nat}{\vnu > 0} \ctcomma y{:}\urt{\Nat}{\top}
  \ctsubseteq{\{y\}} y{:}\urt{\Nat}{\vnu > 0} \ctcomma z{:}\urt{\Nat}{\top}
\end{align*}
\cbnewadding{The direction of the relation is deliberately opposite to
  ordinary type subtyping because typing contexts occur as
  assumptions: the semantic relation ensures that every capability
  satisfying $\Gamma_1$ can be matched on $X$ by one satisfying
  $\Gamma_2$.  Hence, a typing derivation under $\Gamma_2$ can be
  reused under $\Gamma_1$ by applying \textsc{T-CtxSub}.}

\begin{example}[Substructural rules] We can derive the following
  standard substructural rules from \textsc{T-CtxSub}:
  \begin{align*}
    \mprooftr{25mm}
      {$\Gamma_1 \vdash e : \tau$}
      {T-Frame}
      {$\Gamma_1 \ctstar \Gamma_2 \vdash e : \tau$}
    \quad
    \mprooftr{26mm}
      {$\Gamma_1 \vdash e : \tau$}
      {T-Weakening}
      {$\Gamma_1 \ctcomma \Gamma_2 \vdash e : \tau$}
    \quad
    \mprooftr{25mm}
      {$\Gamma_2 \ctstar \Gamma_1 \vdash e : \tau$}
      {T-Exchange}
      {$\Gamma_1 \ctstar \Gamma_2 \vdash e : \tau$}
    \end{align*}\noindent
    The variable-set constraint in the context subtyping relation is
    important.  For example, the framing rule \textsc{T-Frame}
    actually requires that
    $\Gamma_1 \ctstar \Gamma_2 \ctsubseteq{\Code{FV}(e) \cup
      \Code{FV}(\tau)} \Gamma_1$ holds, which is silently enforced by
    $\Gamma_1 \vdash e : \tau$, requiring that the domain of
    $\Gamma_1$ is a superset of $\Code{FV}(e) \cup \Code{FV}(\tau)$.
\end{example}

Similar to other refinement type systems, we require all well-typed
terms to be total. \cbnewadding{Note that an angelic type does not say
  anything about whether a term diverges: a term of type
  $\urt{nat}{\vnu < 2}$ may diverge, as long as it can also evaluate
  to $\Code{0}$ and $\Code{1}$. This is consistent with our
  interpretation of of angelic types as underapproximations of program
  behavior.\footnote{While we could extend our type system so that
    angelic types could explicitly account for diverging behaviors,
    doing so would complicate the system in a way that is orthogonal
    to our treatment of safety and reachability.}}
Thus, when typing the body of a recursive function, \textsc{T-Fix}
uses a demonic refinement to require that the first argument of any
recursive application decreases according to some well-founded
relation $\prec$. As a consequence, the first parameter of a recursive
function needs to be a base type and cannot have an angelic
refinement.  To see why, observe that if we relaxed this restriction,
the recursive argument $v_{\Code{arg}}$ could have the following
type: %
\begin{align*}
  x{:}\urt{b}{\phi} \vdash v_{\Code{arg}} : \urt{b}{\phi} \interty
  \ort{b}{\vnu \prec x}
\end{align*}%
\noindent The first component of this intersection type requires
$v_{\Code{arg}}$ to reach everything that $x$ can, but the second
component prohibits $v_{\Code{arg}}$ from reaching the value of $x$
itself! Intuitively, such a $v_{\Code{arg}}$ cannot exist, meaning
that a function that does not satisfy this restriction can never call
itself recursively.

The totality requirement also forces \textsc{T-Fix} to stipulate that
the first parameter of a recursive function use the entangled arrow
type. When typing the body of the function, the binding for the
recursive call references the original recursive parameter $x$ via the
well-foundedness constraint $\vnu \prec x$, leading to a dependency
between any recursive argument and $x$.
This dependency makes the independence requirement of $\ctwand$
impossible to satisfy: $\ctwand$ requires the argument context to be
independent of the closure context, but well-founded induction
inherently requires the opposite. Notably, although \textsc{T-Fix}
restricts the function type of recursive calls to use $\ctsarr$, the
subsumption rule \textsc{T-Sub} can be used to relax this restriction
afterwards, provided the subtyping relation holds.

  \begin{example}[Fixpoint Function] The following recursive function
    defines Peano subtraction.
\begin{align*}
    \vdash &\zzfix{\Code{sub}}{x}{\zzlam{y}{\zif{x = 0}{0}{\zif{y = 0}{x}{\Code{sub}\;(x - 1)\;(y - 1)}}}}
    \\& : x{:}\ort{\Nat}{\top} \ctsarr y{:}\urt{\Nat}{\vnu \leq x} \ctsarr \urt{\Nat}{\vnu \leq x}
  \end{align*}\noindent
  According to \textsc{T-Fix}, we need to show that
\begin{align*}
  &x{:}\ort{\Nat}{\top} \ctcomma \Code{sub}{:} (x{:}\ort{\Nat}{\vnu \prec x} \ctsarr y{:}\urt{\Nat}{\vnu \leq x} \ctsarr \urt{\Nat}{\vnu \leq x})  \vdash
  \\ &\quad \zzlam{y}{\zif{x = 0}{0}{\zif{y = 0}{x}{\Code{sub}\;(x - 1)\;(y - 1)}}}
    \\&\quad  : y{:}\urt{\Nat}{\vnu \leq x} \ctsarr \urt{\Nat}{\vnu \leq x}
\end{align*}\noindent
When $x = 0$, the program always returns $0$; when $x > 0$, $x - 1$ is a valid argument for the type of $\Code{sub}$, with the well-foundedness constraint $\vnu \prec x$ satisfied. Moreover, when $y$ can cover all values less than or equal to $x$, $y - 1$ can cover all values less than or equal to $x - 1$.
  \end{example}

  Uses of datatype constructors and built-in operations are typed
  using the \textsc{T-AppOp} rule, which is similar to the function
  application rule but relies on the operator typing function
  $\Code{Ty}$. This function uses the intersection of an angelic and
  demonic refinement to assign a precise (i.e., singleton) type to
  operations: %
\begin{align*}
  \Code{Ty}(\Code{O}) &= \ort{\Nat}{\vnu = 0} \interty \urt{\Nat}{\vnu = 0} \\
  \Code{Ty}(\Code{S}) &= n{:}\ort{\Nat}{\top} \ctsarr \ort{\Nat}{\vnu = n+1} \interty \urt{\Nat}{\vnu = n + 1} \\
  \Code{Ty}(+) &= x{:}\ort{\Nat}{\top} \ctsarr y{:}\ort{\Nat}{\top} \ctsarr \ort{\Nat}{\vnu = x+y} \interty \urt{\Nat}{\vnu = x + y}
\end{align*}
\cbnewadding{Datatype constructors are similarly equipped with precise
  types. As an example, suppose that \Code{Cons} has the result
  qualifier $\Code{head}(\vnu,h) \land \Code{tail}(\vnu,t)$ for
  parameters $h$ and $t$. Applying it to $1$ and $[1;2]$ substitutes
  these values for the parameters, yielding
  $\Code{head}(\vnu,1) \land \Code{tail}(\vnu,[1;2])$, which precisely
  characterizes the list $[1;1;2]$.} These sorts of precise types
allow the rule \textsc{T-AppOp} to forward-infer constraints from
operator application results.

The rule for match-expressions requires the target type context to be
divided into a sum of contexts (e.g., using \textsc{T-CtxSub}) that
correspond to the \emph{reachable} control flows under which
individual branches will be executed; the type of the entire
expression is the sum of the types of these branches.
\cbnewadding{\textsc{T-Match} uses a pair of (disjoint) index sets $I$
  and $J$ to identify the reachable and unreachable branches. For
  every $i \in I$, the rule requires a) a type context, $\Gamma_i$,
  that provides types for the pattern variables $\overline{y_i}$ of
  the $i$th branch, b) that $d_i\;\overline{y_i}$ \emph{precisely}
  matches the discriminee $v$ under $\Gamma_i$, and c) that the
  corresponding branch body $e_i$ can also be typed under $\Gamma_i$.
  Conversely, for every $j \in J$, the final premise of
  \textsc{T-Match} establishes that the remaining constructor cannot
  match $v$.}

\begin{example}[Pattern matching] The first two sets of assumptions of
  \textsc{T-Match} deal with the reachable branches of a
  $\mykw{match}$ statement; the last set deals with any branches that
  are unreachable under the current typing context. To illustrate how
  the first two assumptions work, consider the typing of
  \ref{ex:oplusWeak} from the previous section, once we have desugared
  its $\mykw{if}$ expression.
\begin{align*}
  x{:}\urt{\Nat}{\top} \vdash \zmatchNat{x}{3}{y}{5} : \ort{\Nat}{\vnu < 4} \ctoplus \ort{\Nat}{\top}
\end{align*}
Here, we want to show that the angelic context $x{:}\urt{\Nat}{\top}$
ensures that it is possible for this term to reach some value less
than $\Code{4}$. Both branches in \ref{ex:oplusWeak} are reachable, so
we first partition the type context using \textsc{T-CtxSub} into
contexts covering each branch: \cbnewadding{Since this is a
  declarative type system, we can freely manifest the precise
  intermediate contexts needed by each branch via the context
  refinement relation; they need not be inferred directly from the
  syntax of \ref{ex:oplusWeak}.  \footnote{To ensure that these typing
    contexts can be combined via $\ctoplus$, each $\Gamma_i$ must also
    provide bindings for the pattern variables used by the other
    reachable branches, even though they are not used by the $i$th
    branch.  These variables can simply be assigned types with the top
    qualifier, as $\Gamma_{\Code{TBranch}}$ does for the unused
    pattern variable $y$.}} %
{\small %
  \begin{align*}
    &x{:}\urt{\Nat}{\top} \ctsubseteq{\{x\}} \\
    & \;\;
      \underbrace{x{:}\ort{\Nat}{\vnu = 0}\interty \urt{\Nat}{\vnu =
      0} \ctcomma
      y{:}\ort{\Nat}{\top} }_{\Gamma_{\Code{TBranch}}} \ctoplus
      \underbrace{x{:}\ort{\Nat}{\vnu > 0}\interty
      \urt{\Nat}{\vnu > 0} \ctcomma
      y{:}\ort{\Nat}{\vnu = x - 1} }_{\Gamma_{\Code{FBranch}}}
  \end{align*}}\noindent
These contexts satisfy the first set of assumptions of
\textsc{T-Match}:
\begin{align*}
  &\Gamma_{\Code{TBranch}} \vdash \Code{O} :
  \ort{\Nat}{\vnu = 0} \interty \urt{\Nat}{\vnu = 0} \\
  &\Gamma_{\Code{FBranch}} \vdash \Code{S}\;y :
  \ort{\Nat}{\vnu = x} \interty \urt{\Nat}{\vnu = x}
\end{align*}
\noindent
We can then type each branch under these contexts to show that
\ref{ex:oplusWeak} has the target type (since we have accounted for
both branches, the third set of assumptions trivially holds):\smallskip
\begin{center}
  \begin{prooftree}
    \hypo{\Gamma_{\Code{TBranch}} \vdash 3 : \ort{\Nat}{\vnu < 4}}
    \hypo{\Gamma_{\Code{FBranch}} \vdash 5 : \ort{\Nat}{\top}}
    \infer2[\textsc{T-Match}]{\Gamma_{\Code{TBranch}} \ctoplus \Gamma_{\Code{FBranch}}
      \vdash  \zmatchNat{x}{3}{y}{5} : \ort{\Nat}{\vnu < 4} \ctoplus
      \ort{\Nat}{\top}}
    \end{prooftree}
\end{center}\smallskip\noindent
Observe we are unable to check \ref{ex:oplusWeak} against this type
under a weaker typing context:
\begin{align*}
  x{:}\urt{\Nat}{\vnu > 0} \vdash \zmatchNat{x}{3}{y}{5} :
  \ort{\Nat}{\vnu < 4} \ctoplus \ort{\Nat}{\top} \quad \judgfail
\end{align*}\noindent
Here, $x{:}\urt{\Nat}{\vnu > 0}$ admits a capability containing
environments that only reach the false branch, and thus cannot be
decomposed into two contexts that fully separate both branches.  The
following example illustrates how \textsc{T-Match}'s last set of
assumptions accounts for unreachable control flows:
\begin{align*}
  x{:}\ort{\Nat}{\vnu > 0} \ctcomma y{:}\ort{\Nat}{\vnu = x - 1} \vdash \zmatchNat{x}{3}{y}{5} : \ort{\Nat}{\vnu > 4}
\end{align*}\noindent
By \textsc{T-CtxSub} and the fact that
\begin{align*}
  x{:}\ort{\Nat}{\vnu > 0} \ctsubseteq{\{x\}}
      x{:}\ort{\Nat}{\vnu > 0} \ctcomma y{:}\ort{\Nat}{\vnu = x - 1}
\end{align*}\noindent
it suffices to establish the following derivation using
\textsc{T-Match}:
\smallskip
\begin{center}
  \small
  \begin{prooftree}
    \hypo{x{:}\ort{\Nat}{\vnu > 0} \ctcomma y{:}\ort{\Nat}{\vnu = x - 1}\vdash 5 : \ort{\Nat}{\vnu > 4}}
    \infer[no rule]1{x{:}\ort{\Nat}{\vnu > 0} \ctcomma y{:}\ort{\Nat}{\vnu = x - 1}\vdash \Code{S}\;y : \ort{\Nat}{\vnu = x} \interty \urt{\Nat}{\vnu = x}}
    \infer[no rule]1{x{:}\ort{\Nat}{\vnu > 0} \ctcomma y{:}\ort{\Nat}{\vnu = x - 1} \vdash \Code{O} : \ort{\Nat}{\vnu \neq x}}
    \infer1[\textsc{T-Match}]{x{:}\ort{\Nat}{\vnu > 0} \ctcomma y{:}\ort{\Nat}{\vnu = x - 1} \vdash \zmatchNat{x}{3}{y}{5} : \ort{\Nat}{\vnu > 4} \quad \judgok}
    \end{prooftree}
  \end{center}\noindent
\end{example}

\section{Metatheory} %
\label{sec:meta}

In this section, we provide the denotational semantics of types
($\denot{\tau}$) and type contexts ($\denot{\Gamma}$), and state the
soundness theorem of our type system.  The key challenge is to give a
semantics that supports both demonic and angelic modalities
simultaneously, together with the resource-based interpretation of
contexts that tracks variable dependencies.  We proceed in four
steps: we define contextual capabilities as the semantic domain
(Section~\ref{subsec:capabilities}), equip them with algebraic structure
(Section~\ref{subsec:algebra}), build a context logic over this algebra
(Section~\ref{subsec:logic}), and finally give the type denotation and
soundness theorem (Section~\ref{subsec:denotation}).

\subsection{Contextual Capabilities}%
\label{subsec:capabilities}

In a pure functional language, the free variables of a term $e$ are
bound by its evaluation environment, a partial map from variables to
values.  To reason about both safety and reachability, we need to
consider not just a single environment but a \emph{set} of
environments, representing the range of possible variable bindings
available to demonic or angelic choices.  We call such a set a
\emph{contextual capability} or capability for short.

\begin{definition}[Contextual Capability]\label{def:contextual-capability}
  A \emph{contextual capability} is a non-empty set of environments
  (partial maps, $\Var \rightharpoonup \Val$) that all share the same
  domain.  We use $r$ to range over capabilities and $\Res$ as the set
  of all capabilities.
\end{definition}

The domain condition ensures that all environments in a capability
bind exactly the same set of variables, reflecting the fact that the
free variables of a term are fixed by its typing context.  A
capability with a single environment corresponds to a fully determined
variable binding; a capability with multiple environments encodes the
range of choices available when reasoning about safety or
reachability.

\cbnewadding{We write $\Dom{r}$ for the domain shared by all
  environments in $r$.  The projection $\projectTo{r}{X}$ restricts
  every environment in $r$ to $X \cap \Dom{r}$.  Two capabilities are
  compatible, written $r_1 \uparrow r_2$, if every pair of
  environments drawn from them agrees on the variables shared by their
  domains.  Finally, $r_1 \sqsubseteq r_2$ holds when
  $r_1 = \projectTo{r_2}{\Dom{r_1}}$.} The refinement relation
$\sqsubseteq$ plays a central role: $r_1 \sqsubseteq r_2$ says that
$r_1$ is a restriction of $r_2$ to a smaller variable domain, i.e.,
$r_2$ extends the bindings of $r_1$ with additional variables.  This
relation will serve as the Kripke accessibility relation in our logic.

To express variable dependency in refinement type qualifiers, we also
need the following notion.

\begin{definition}[Fiber] %
  \label{def:fibers}
  For a variable $x$ and an environment $[x \mapsto v] \in
  \projectTo{r}{\{x\}}$, the \emph{fiber} of $r$ over $[x \mapsto
  v]$ is:
  $
    \fiberOver{r}{[x \mapsto v]} \doteq \{\, \sigma \in r \mid
    \sigma(x) = v \,\}.
  $
  That is, $\fiberOver{r}{[x \mapsto v]}$ is the largest subset of $r$
  comprised of environments that map $x$ to $v$.
\end{definition}
\noindent Fibers allow us to condition a capability on the value of a
specific variable, which is essential for evaluating refinement
qualifiers that refer to other variables bound in the typing context
(see Section~\ref{subsec:logic}).

\subsection{Capability Algebra} %
\label{subsec:algebra}

Our context logic is a bunched implication (BI) logic~\cite{OHP99}
based on a BI-algebra~\cite{Pym+02}, a partial commutative monoid with
a pre-order satisfying bifunctoriality.  We equip contextual
capabilities with two operations: a \emph{product} that composes
disjoint variable bindings, and a \emph{sum} that merges capabilities
over the same domain to support control-flow reasoning. The
\emph{product} $r_1 \times r_2$ operation, defined when
$r_1 \uparrow r_2$, is the set of all pairwise unions
$\sigma_1 \cup \sigma_2$ with $\sigma_1 \in r_1$ and
$\sigma_2 \in r_2$.
The \emph{sum} operation, $r_1 + r_2$, defined when
$\Dom{r_1} = \Dom{r_2}$, is simply $r_1 \cup r_2$. Together with unit
$\mathbf{1} \doteq \{\emptyset\}$, the tuple
$(\Res, +, \times, \mathbf{1})$ forms a partial commutative semiring
without zero.

To enable a Kripke semantics for BI-logic, this semiring must be
equipped with a partial order satisfying bifunctoriality.  The natural
candidates for this order are the subset ($\subseteq$) and superset
($\supseteq$) relations.  However, neither works for both angelic and
demonic modalities simultaneously.  \cbnewadding{To see why, consider
  the following two possible accessibility steps from $r_1$:}
\begin{equation}
\label{eq:kripke-neutrality}
\begin{array}{c @ {\hspace{1em}} c @ {\hspace{1em}} l}
  \multirow{2}{*}{$\boxed{r_1 \doteq \{[x \mapsto 1], [x \mapsto 2]\}}$}
  & \subseteq
  & \boxed{r_2 \doteq \{[x \mapsto 1], [x \mapsto 2], [x \mapsto 3]\}} \\[0.5em]
  & \sqsubseteq
  & \boxed{r_3 \doteq \{[x \mapsto 1; y \mapsto 3], [x \mapsto 2; y \mapsto 4]\}}
\end{array}
\end{equation}
\cbnewadding{The predicate $1 \leq x \leq 2$ clearly holds for both
  environments in $r_1$.  Consider the superset $r_2$.  Under an
  angelic reading of this predicate, both $1$ and $2$ must be
  available choices for $x$, a condition satisfied by $r_2$.  Under
  the demonic reading, every possible value of $x$ must lie between
  $1$ and $2$, so the inclusion of $x \mapsto 3$ in $r_2$ means it
  does not satisfy the demonic interpretation.  Intuitively, adopting
  $\subseteq$ as our accessibility order would favor an angelic
  reading over a demonic one.  A similar argument applies to
  $\supseteq$. In contrast, the environment $r_3$ adds a binding for
  $y$ while precisely preserving the available choices for $x$, and is
  neutral with respect to both readings: the predicate continues to
  hold under both interpretations.}  As this example illustrates, the superset order supports an
overapproximation semantics (demonic), while the subset order supports
underapproximation (angelic).  Since we need to support both as
first-class operators, the Kripke order must be \emph{neutral} with
respect to approximation.  The refinement relation $\sqsubseteq$ has
exactly this property: $r \sqsubseteq r'$ holds if $r'$ extends $r$
with additional bindings, while preserving all existing environments,
neither adding nor removing possible values for the variables already
in $r$'s domain.  Specifically, $\sqsubseteq$ satisfies the standard
bifunctoriality property required of a BI-algebra:
$r_1 \sqsubseteq r_1'$ and $r_2 \sqsubseteq r_2'$ implies
$r_1 \times r_2 \sqsubseteq r_1' \times r_2'$, ensuring that
$(\Res, +, \times, \mathbf{1}, \sqsubseteq)$ forms a BI-algebra.

\subsection{Context Logic} %
\label{subsec:logic}

We now define a logic over our capability algebra.  The logic is a BI
logic extended with demonic and angelic modalities, a sum operator for
control-flow reasoning, and a \emph{binding reference operator} that
expresses how a qualifier may depend on specific variable bindings in
the current capability.

\begin{definition}[Context Logic]\label{def:context-logic}
  The syntax of our context logic is:
  \begin{align*}
    P, Q \;::=\; &\chtrue \mid \chfalse \mid P \chland Q \mid P \chlor Q \mid P \chimpl Q \mid P \chstar Q \mid P \chwand Q
    \\& \mid \chforall x.\, P \mid \chexists x.\, P \mid \Code{Atom}(\phi)
    \mid \demonic P \mid \angelic P \mid P \choplus Q \mid \forallM{x} P
  \end{align*}
  The forcing semantics over capabilities $r \in \Res$ is:
  \begin{itemize}
  \item $r \models \chtrue$ always; \quad $r \models \chfalse$ never;
  \item $r \models P \chland Q$ iff $r \models P$ and $r \models Q$;
  \item $r \models P \chlor Q$ iff $r \models P$ or $r \models Q$;
  \item $r \models P \chimpl Q$ iff for all $r'$ such that $r \sqsubseteq r'$, $r' \models P$ implies $r' \models Q$;
  \item $r \models P \chstar Q$ iff $\exists r_1, r_2.\; r_1 \times r_2 \sqsubseteq r$,\; $r_1 \models P$,\; $r_2 \models Q$;
  \item $r \models P \chwand Q$ iff $\forall r' \uparrow r$,\; $r' \models P$ implies $r' \times r \models Q$;
  \item $r \models \chforall x.\, P$ iff $x \notin \Dom{r}$ and $\forall r'$ such that $r \sqsubseteq r'$ with $x \in \Dom{r'}$,\; $r' \models P$;
  \item $r \models \chexists x.\, P$ iff $x \notin \Dom{r}$ and $\exists r'$ such that $r \sqsubseteq r'$ with $x \in \Dom{r'}$,\; $r' \models P$;
  \item $r \models \Code{Atom}(\phi)$ iff $\projectTo{r}{\Code{FV}(\phi)} = \{\,\sigma \mid \Dom{\sigma} = \Code{FV}(\phi) \land \sigma \models \phi\,\}$;
  \item $r \models \demonic P$ iff $\exists r', r \subseteq r'$ such that $r' \models P$;
  \item $r \models \angelic P$ iff $\exists r', r' \subseteq r$ such that $r' \models P$;
  \item $r \models P \choplus Q$ iff $\exists r_1, r_2.\; r_1 + r_2 \sqsubseteq r$,\; $r_1 \models P$,\; $r_2 \models Q$;
  \item $r \models \forallM{x} P$ iff $\forall [x \mapsto v] \in \projectTo{r}{\{x\}}.\; \fiberOver{r}{[x \mapsto v]} \models P[x \mapsto v]$.\footnote{
We extend the binding operator to sets of variables by iterated
application: for a finite set $X = \{x_1, \ldots, x_n\}$,
we write $\forallM{X}\, P$ as shorthand
for
$\forallM{x_1}\,\forallM{x_2}\,\cdots\,\forallM{x_n}\,P$
In particular,
$\forallM{\Dom{\Sigma}}\, P$ binds all
variables in the domain of the current basic type context $\Sigma$
before evaluating $P$.}
  \end{itemize}
  We write $P \models Q$ iff for all $r \in \Res$, $r \models P$ implies $r \models Q$.
\end{definition}

The additive connectives ($\chland$, $\chlor$, $\chimpl$, $\chforall$,
$\chexists$) are standard.  The multiplicative connectives ($\chstar$,
$\chwand$) follow standard BI-logic semantics using our algebra's
product.  We discuss the remaining operators below.

\paragraph{Kripke Monotonicity}
The implication $P \chimpl Q$ and the first-order quantifiers
$\chforall x.\, P$ and $\chexists x.\, P$ use the refinement relation
$\sqsubseteq$ as the Kripke accessibility relation: truth at $r$
requires truth at all $r'$ such that $r \sqsubseteq r'$, i.e., at all
extensions of $r$ with additional variable bindings.  This is the
standard Kripke semantics for implication in BI-logic, and ensures the
following monotonicity property.

\begin{theorem}[Kripke Monotonicity]\label{thm:kripke-monotonicity}
  $r \sqsubseteq r'$ and $r \models P$ implies $r' \models P$.
\end{theorem}

The use of $\sqsubseteq$ rather than $\subseteq$ or $\supseteq$ as the
accessibility relation is what allows our logic to remain neutral with
respect to approximation direction, per \autoref{subsec:algebra}.

\paragraph{Atomic Propositions}
The atomic proposition $\Code{Atom}(\phi)$ lifts a standard formula
$\phi$ to a capability-level predicate with \emph{exact} semantics: a
capability $r$ satisfies $\Code{Atom}(\phi)$ iff its projection to
$\Code{FV}(\phi)$ contains \emph{exactly} those environments that
satisfy $\phi$.  This neutrality leaves the choice of approximation
direction entirely to the demonic and angelic modalities.

\paragraph{Demonic and Angelic Modalities}
The demonic modality $\demonic P$ holds at $r$ if there exists an
upper bound $r'$ (i.e., $r \subseteq r'$) that satisfies $P$; this
models overapproximation since $r'$ extends $r$ with additional
environments.  Dually, the angelic modality $\angelic P$ holds at $r$
if some lower bound $r' \subseteq r$ satisfies $P$; this models
underapproximation since $r'$ witnesses a specific subset of the
environments in $r$.
\begin{example}%
  \label{ex-approximation-modalities}
  Let $r_x$ be $\{[x \mapsto 1], [x \mapsto 2]\}$. The following hold:
  \begin{align*}
    &r_x \models \demonic\,\Code{Atom}(x = 1) \;\judgfail \quad
    &&r_x \models \demonic\,\Code{Atom}(0 \leq x \leq 4) \;\judgok \\
    &r_x \models \angelic\,\Code{Atom}(x = 1) \;\judgok \quad
    &&r_x \models \angelic\,\Code{Atom}(0 \leq x \leq 4) \;\judgfail
  \end{align*}
  The demonic assertion $\demonic\,\Code{Atom}(x = 1)$ fails because
  no superset of $r_x$ satisfies $\Code{Atom}(x = 1)$ exactly (since
  $r_x$ already contains $[x \mapsto 2]$, any superset violates
  exactness).  The angelic assertion
  $\angelic\,\Code{Atom}(0 \leq x \leq 4)$ fails because no subset of
  $r_x$ satisfies $\Code{Atom}(0 \leq x \leq 4)$ exactly (that would
  require environments for $x = 0, 3, 4$ which $r_x$ does not
  contain).

  Critically, satisfiability of demonic and angelic modalities is also
  maintained by refinement.  Consider the following capabilities:
\begin{align*}
\begin{array}{c @{\hspace{1.5em}} c @{\hspace{1.5em}} c}
  {\boxed{r_x \doteq \{[x \mapsto 1], [x \mapsto 2]\}}}
  & \sqsubseteq
  & {\boxed{r_2 \doteq \{[x \mapsto 1; y \mapsto 2], [x \mapsto 2; y \mapsto 7]\}}} \\[0.2em]
  \vsubseteq & & \vsubseteq \\[0.2em]
  {\boxed{r_3 \doteq \{[x \mapsto 1]\}}}
  & \sqsubseteq
  & {\boxed{r_4 \doteq \{[x \mapsto 1; y \mapsto 2]\}}}
\end{array}
\end{align*}\noindent
For an angelic formula $\angelic P$, $r_3 \models \angelic P$ implies
$r_4 \models \angelic P$ by horizontal refinement, and then
$r_2 \models \angelic P$ by the vertical subset relation.  Dually,
$r_x \models \demonic P$ implies $r_2 \models \demonic P$ by
horizontal refinement, and then $r_4 \models \demonic P$ by the
vertical subset relation.  Thus, the direction of the subset relation
is reversed between the two modalities, but the direction of
refinement is unchanged.

Finally, to illustrate how the product operation interacts with the
logic's two conjunction operators, let
$r_y \doteq \{[y \mapsto 3], [y \mapsto 4]\}$ and
$r_{xy} \doteq \{[x \mapsto 1, y \mapsto 3], [x \mapsto 2, y \mapsto
4]\}$.  Then:
\begin{align*}
&r_x \times r_y \models (\angelic\,\Code{Atom}(1 \leq x \leq 2)) \chstar (\angelic\,\Code{Atom}(3 \leq y \leq 4)) \;\judgok \\
&r_{xy} \models (\angelic\,\Code{Atom}(1 \leq x \leq 2)) \chland (\angelic\,\Code{Atom}(3 \leq y \leq 4)) \;\judgok \\
&r_{xy} \models (\angelic\,\Code{Atom}(1 \leq x \leq 2)) \chstar (\angelic\,\Code{Atom}(3 \leq y \leq 4)) \;\judgfail
\end{align*}
The product $r_x \times r_y$ satisfies the multiplicative conjunction
because $x$ and $y$ are independently bound.  The capability $r_{xy}$
satisfies the additive conjunction but not the multiplicative one,
since its bindings are entangled and cannot be decomposed into two
disjoint sub-capabilities.

\end{example}

\paragraph{Sum Operator}
The sum $P \choplus Q$ holds at $r$ if $r$ can be split into two
sub-capabilities $r_1 + r_2 \sqsubseteq r$, one satisfying $P$ and one
satisfying $Q$.  This supports the context sum type
$\tau_1 \ctoplus \tau_2$ from \autoref{sec:context-typing}, which
allows a capability to be partitioned into two subsets. The requirement
that capabilities be non-empty ensures that neither branch is vacuous:
both summands must contain at least one environment.

\paragraph{Binding Reference Operator}
The binding reference operator $\forallM{x} P$ expresses how a
qualifier may depend on the specific binding of variable $x$ in the
current capability.  It holds at $r$ if, for every possible binding
$v$ of $x$ in $r$, the sub-capability $\fiberOver{r}{[x \mapsto v]}$
satisfies $P$ with $x$ substituted by $v$.  This is needed because
$\Code{Atom}(\phi)$ has exact semantics: a qualifier $\phi$ mentioning
$x$ requires a satisfying capability $r$ to contain \emph{all}
environments consistent with $x$'s full range, which may conflict with
the constrained range of $x$ expressed by other conjuncts.  The
binding reference operator resolves this by fixing the binding of $x$
before evaluating the qualifier, enabling dependent qualifiers to
refer to specific variable bindings in the capability.

\begin{example}%
  \label{ex-variable-reference}
  Consider the entangled context
  \[x{:}\urt{\Nat}{1 \leq \vnu
    \leq 2} \ctcomma y{:}\urt{\Nat}{0 \leq \vnu < x}\]
\noindent where $y$'s
  range depends on $x$.  Its denotation is:
  \[
    \denot{x{:}\urt{\Nat}{1 \leq \vnu \leq 2} \ctcomma y{:}\urt{\Nat}{0 \leq \vnu < x}}
    = \angelic\,(1 \leq x \leq 2) \chland \forallM{x}\,\angelic\,(0 \leq y < x).
  \]
  The operation $\forallM{x}\,\angelic\,(0 \leq y < x)$ fixes the
  binding of $x$ before evaluating $\angelic\,(0 \leq y < x)$, so
  $y$'s required range is determined separately for each value of $x$.
  Capabilities that model this formula include
  $\{[x \mapsto 1;\, y \mapsto 0]\}$ and
  $\{[x \mapsto 2;\, y \mapsto 0];\, [x \mapsto 2;\, y \mapsto 1]\}$,
  as well as their sum. Omitting $\forallM{x}$ from the right conjunct
  would require a satisfying capability to include an environment with
  a mapping for \emph{every} pair of $x$ and $y$ where $0 \leq y < x$:
  $[x \mapsto 3;\, y \mapsto 0], [x \mapsto 4;\, y \mapsto 1],
  \ldots$
\end{example}

\subsection{Type Denotation and Soundness}%
\label{subsec:denotation}

We now give the denotation of types and type contexts in terms of the
context logic defined above, and state the main soundness theorem.  We
write $\denot{\tau}_\Sigma : \Code{Term} \to \Code{Formula}$ for the
type denotation under a basic type context $\Sigma$, and
$r \models \denot{\tau}_\Sigma\;e$ to mean that $e$ inhabits type
$\tau$ under capability $r$.  We omit $\Sigma$ when it is clear from
context.

As is customary in other refinement types, we require terms to be
total and well-typed in the basic type system.  We encode these side
conditions into the logic via a guard formula:
\begin{align*}
  P_{\Code{guard}} \doteq \forallM{\Dom{\Sigma}}\,\Code{Atom}(\Code{Total}(e) \land \Sigma \vdash_s e : \eraserf{\tau} \land \Sigma \vdash_{\Code{WF}} \tau).
\end{align*}
The binding reference operator $\forallM{\Dom{\Sigma}}$ is necessary
here; without it, e.g., $\Code{Atom}(\Sigma \vdash_s x : \Nat)$ is
only satisfied by a capability that contains an environment for every
possible value of $x$.  All type denotations below implicitly include
this guard, which we omit for brevity.

The denotation of a term $e$ under capability $r$ captures the image
of $e$ under all environments in $r$ via a variable $x$:
\begin{align*}
  \denot{\mstep{e}{x}}_\Sigma &\doteq \forallM{\Dom{\Sigma}}\,\Code{Atom}(\mstep{e}{x}).
\end{align*}
The binding reference operator $\forallM{\Dom{\Sigma}}$ fixes the
bindings of all variables in the typing context before evaluating the
atomic reduction predicate, ensuring that the reduction $\mstep{e}{x}$
is checked under each specific environment in the current capability
rather than requiring the capability to contain all environments
consistent with $x$'s full range.

\begin{definition}[Type Denotation]\label{def:type-denotation}
  \begin{align*}
    \denot{\ort{b}{\phi}}_\Sigma\;e &\doteq \denot{\mstep{e}{\vnu}}_\Sigma \chimpl \forallM{(\Code{FV}(\phi)-\{\vnu\})}\,\demonic\,\Code{Atom}(\phi) \quad \text{($\vnu$ fresh)}  \\
    \denot{\urt{b}{\phi}}_\Sigma\;e &\doteq \denot{\mstep{e}{\vnu}}_\Sigma \chimpl \forallM{(\Code{FV}(\phi)-\{\vnu\})}\,\angelic\,\Code{Atom}(\phi) \quad \text{($\vnu$ fresh)}  \\
    \denot{x{:}\tau_x \ctsarr \tau}_\Sigma\;e &\doteq \denot{\tau_x}_\Sigma\;x \chimpl \denot{\tau}_{\Sigma,x{:}\eraserf{\tau_x}}\;(e\;x) \quad \text{($x$ fresh)}  \\
    \denot{x{:}\tau_x \ctwand \tau}_\Sigma\;e &\doteq \denot{\tau_x}_\Sigma\;x \chwand \denot{\tau}_{\Sigma,x{:}\eraserf{\tau_x}}\;(e\;x) \quad \text{($x$ fresh)}  \\
    \denot{\tau_1 \ctinter \tau_2}_\Sigma\;e &\doteq \denot{\tau_1}_\Sigma\;e \chland \denot{\tau_2}_\Sigma\;e  \\
    \denot{\tau_1 \ctoplus \tau_2}_\Sigma\;e &\doteq \denot{\tau_1}_\Sigma\;e \choplus \denot{\tau_2}_\Sigma\;e
  \end{align*}
\end{definition}

Demonic and angelic type denotations follow directly from the
corresponding modalities: the result of evaluating $e$ bound to
fresh variable $\vnu$ must satisfy $\phi$ under either the demonic or
angelic interpretation, with the binding operator fixing the bindings
of the free variables of $\phi$ (besides $\vnu$) before applying the modality.  The
function type denotations internalize the adjunction between function
types and context combinators: $\ctsarr$ uses additive implication
(the parameter binding is entangled with the closure context), while
$\ctwand$ uses multiplicative implication (the parameter binding is
disjoint from the closure context).

\begin{definition}[Type Context Denotation]\label{def:context-denotation}
  \begin{align*}
    \denot{\emptyset}_\Sigma & \doteq \chtrue \quad
    & \denot{\Gamma_1 \ctcomma \Gamma_2}_\Sigma & \doteq \denot{\Gamma_1}_\Sigma \chland \denot{\Gamma_2}_{(\Sigma,\,\eraserf{\Gamma_1})}
                             & \denot{\Gamma_1 \ctoplus \Gamma_2}_\Sigma & \doteq \denot{\Gamma_1}_\Sigma \choplus \denot{\Gamma_2}_\Sigma \\
    \denot{x{:}\tau}_\Sigma \doteq & \denot{\tau}_\Sigma\;x
    & \denot{\Gamma_1 \ctstar \Gamma_2}_\Sigma & \doteq \denot{\Gamma_1}_\Sigma \chstar \denot{\Gamma_2}_\Sigma
  \end{align*}
\end{definition}

Entangled conjunction ``$\ctcomma$'' maps to additive conjunction
$\chland$, reflecting that the second context may refer to bindings in
the first.  Independent conjunction ``$\ctstar$'' maps to
multiplicative conjunction $\chstar$, requiring the two
sub-capabilities to be decomposable as a product.

\begin{figure}[h]
  {
  \begin{flalign*}
    &\text{\textbf{Semantic subtyping}} && \fbox{$\Gamma \vdash \tau \subty \tau$ \quad $\Gamma \ctsubseteq{X} \Gamma$}
  \end{flalign*}
  \mprooftr{42mm}
  {$\forall e.\; \denot{\Gamma} \models \denot{\tau_1}\;e \chimpl \denot{\tau_2}\;e$}
  {Sub}
  {$\Gamma \vdash \tau_1 <: \tau_2$}
  \quad
  \mprooftr{66mm}
  {$\forall r.\; r \models \denot{\Gamma_1} \implies \exists r'.\; \projectTo{r}{X} = \projectTo{r'}{X} \land r' \models \denot{\Gamma_2}$}
  {CtxSub}
  {$\Gamma_1 \ctsubseteq{X} \Gamma_2$}
  }
  \caption{Semantic subtyping relations. All denotations use an empty
    basic typing context ($\Sigma = \emptyset$).}
  \label{fig:semantic-subtyping-relation}
\end{figure}

The semantic subtyping relations used in rules \textsc{T-Sub} and
\textsc{T-CtxSub} are shown in
\autoref{fig:semantic-subtyping-relation}. Type subtyping is encoded
directly as logical implication in the denotational semantics.
Context subtyping uses an explicit projection $\projectTo{r}{X}$: the
two capabilities agree on the variable set $X$ (the free variables of
the term and type), but $r'$ may differ on other variables, reflecting
the fact that context subtyping only needs to hold for the variables
that actually appear in the judgement.

We conclude with the core soundness results.

\begin{definition}[Well-Formed Operator Typing Function] %
  \label{thm:well-formed-operator}
  The operator typing function $\Code{Ty}$ is well-formed iff for all primitive
  operators $\primop$: {\small\begin{align*} \Code{Ty}(\primop) =
      \overline{x{:}\ort{b_x}{\phi_x}} \ctsarr \ort{b}{\phi} \ctinter
      \urt{b}{\phi} \quad\text{and}\quad \models
      \denot{\overline{x{:}\ort{b_x}{\phi_x}}} \chiff
      \denot{\ort{b}{\phi} \ctinter
        \urt{b}{\phi}}\;(\primop\;\overline{x}).
  \end{align*}}
\end{definition}

Our fundamental theorem and type soundness theorem assume a
well-formed operator typing function.

\begin{theorem}[Fundamental Theorem]%
  \label{thm:fundamental}
  \langname{}'s typing judgement
  is sound with respect to the denotation of types into context logic:
  $\Gamma \vdash e : \tau$ implies
  $\denot{\Gamma}_\emptyset \models
  \denot{\tau}_{\eraserf{\Gamma}}\;e$. %
\end{theorem}

\begin{corollary}[Type Soundness]%
  \label{thm:soundness}
  $\emptyset \vdash e : \tau$ implies that for any fresh variable $x$,
  $\{[x \mapsto v] \mid \mstep{e}{v}\} \models \denot{\tau}\;x$.
\end{corollary}

Our type soundness theorem simply instantiates the fundamental theorem
with the empty context. For a closed well-typed term $e$, the
capability $\{[x \mapsto v] \mid \mstep{e}{v}\}$, which collects all
values $e$ can reduce to, each bound to a fresh variable $x$,
satisfies the denotation of $\tau$. This has two concrete consequences
depending on the modality of $\tau$: if $\tau$ is a demonic type, then
every value that $e$ reduces to satisfies $\tau$'s qualifier (safety);
if $\tau$ is an angelic type, then every value prescribed by $\tau$'s
qualifier is witnessed by some reduction of $e$ (reachability).

\section{Supporting Nondeterminism}
\label{sec:extension}

The \langname{} language of Section~\ref{sec:context-typing} is
deterministic: every term reduces to (at most) one value under any
particular environment. This section considers a calculus,
\langNDname{}, that relaxes this restriction, enabling us to, e.g.,
type programs like \ref{ex:DisjCon} from
\autoref{sec:overview}. Concretely, \langNDname{} extends \langname{}
with nondeterministic \emph{generators} that can produce multiple
values of basic type across different executions:
\begin{align*}
  \primop ::= \quad & \ldots ~|~ \Code{gen}_b ~|~ \ldots
\end{align*}
\cbnewadding{Using $\Code{gen}$, we can define a nondeterministic
  choice operator
  $e_1 \oplus e_2 \triangleq \zif{\Code{gen_{bool} \;()}}{e_1}{e_2}$.}
Unsurprisingly, the wrinkle introduced by nondeterminism is related to
dependence: whereas the reachability properties of a \langname{} term
were fundamentally tied with the environment under which it is
evaluated, those of a \langNDname{} term may (or may not) be
\emph{independent} of the surrounding context.

Dealing with this new possibility raises two technical
challenges. First, nondeterministic functions like generators can be
invoked multiple times, producing independent capabilities at each
call; our extended type system must account for this \emph{resource
  duplication}.  Second, we need to generalize our type denotation to
accommodate terms whose reachability guarantees are not completely
determined by their evaluation environment.

\subsection{Resource Duplication and Persistence} %
\label{subsec:duplication}

In \langname{}, the reachability properties of a term are entangled
with its surrounding context: evaluation does not introduce new
capabilities. For example, every inhabitant of
$\ort{\Unit}{\top} \ctsarr \urt{\Nat}{\vnu > 0}$ in \langname{} must
capture some variable with an angelic binding that provides the
witnesses needed to construct every positive number:
\begin{align*}
  z{:}\urt{\Nat}{\top} \vdash \zzlam{x}{z + 1} : \ort{\Unit}{\top} \ctsarr
  \urt{\Nat}{0 < \vnu}
\end{align*}\noindent
Two successive calls to a \langname{} function with this signature
\emph{always} yield entangled results, even if the calls have
independent arguments:
\begin{align}
  z{:}\urt{\Nat}{\top} \vdash
  \zlet{f}{\zzlam{x}{z}}{\zlet{y_1}{f\;()}{\zlet{y_2}{f\;()}{y_1 -
  y_2}}} : \urt{\Nat}{\top} \quad \judgfail
  \tag{$\Code{e_{ND}}$}\label{ex:nondet}
\end{align}\noindent
This fact is encoded in the typing rules for let-expressions,
\textsc{T-Let} and \textsc{T-LetI}, which jointly enforce an
``entangled xor consumed'' policy. The rule \textsc{T-Let} allows the
function $f$ to be used in both the bound expression and the body of a
\Code{let}-expression, but the bound expression will be entangled with
$f$. Thus, if we tried to use \textsc{T-Let} to establish
\ref{ex:nondet}, we would get stuck when trying to type
$y_1 - y_2$:

\begin{center}
\small
  \begin{prooftree}
    \hypo{\ldots \ctcomma y_1{:}\urt{\Nat}{\top} \ctcomma
    y_2{:}\urt{\Nat}{\top} \vdash y_1 - y_2 : \urt{\Nat}{\top} \quad \judgfail}
  \infer1[\textsc{T-Let}]{\emptyset \ctcomma z{:}\urt{\Nat}{\top} \ctcomma f{:}\ort{\Unit}{\top}
    \ctsarr \urt{\Nat}{\top} \ctcomma y_1{:}\urt{\Nat}{\top} \vdash
    \zlet{y_2}{f\;()}{y_1 - y_2} : \tau}
\infer1[\textsc{T-Let}]{\emptyset \ctcomma z{:}\urt{\Nat}{\top}
  \ctcomma f{:}\ort{\Unit}{\top} \ctsarr \urt{\Nat}{\top} \vdash
  \zlet{y_1}{f\;()}{\zlet{y_2}{f\;()}{y_1 - y_2}} : \tau}
\end{prooftree}
\end{center} \smallskip

\noindent On the other hand, \textsc{T-LetI} produces an independent
result, but prevents $f$ from reappearing in the body of a
\Code{let}-expression, leaving us unable to type the second
\Code{let}-expression in \ref{ex:nondet}: \smallskip

\begin{center}
\small
  \begin{prooftree}
    \hypo{\emptyset \ctstar y_1{:}\urt{\Nat}{\top} \vdash
      \zlet{y_2}{f\;()}{y_1 - y_2} : \tau \quad \judgfail}
  \infer1[\textsc{T-LetI}]{\emptyset \ctstar (z{:}\urt{\Nat}{\top}
    \ctcomma f{:}\ort{\Unit}{\top} \ctsarr \urt{\Nat}{\top}) \vdash
    \zlet{y_1}{f\;()}{\zlet{y_2}{f\;()}{y_1 - y_2}} : \tau}
  \end{prooftree}
\end{center} \smallskip

Successive applications of a nondeterministic generator like
$\Code{gen_{nat}\; ()}$, however, can produce genuinely independent
values. In order to accurately capture the behavior of such functions,
\langNDname{} introduces a \emph{persistent} type operator, $\ctpers$:
\begin{align*}
  \Code{gen_{nat}} : \ctpers (\ort{\Unit}{\top} \ctsarr \urt{\Nat}{\top})
\end{align*}
Persistence allows a binding in the type context to be duplicated,
e.g.,
\begin{align*}
  x{:}\ctpers \tau \; \ctsubseteq{\{x\}} \; x{:}\ctpers \tau \ctstar x{:}\ctpers \tau
\end{align*}\noindent
When combined with \textsc{T-LetI}, a persistent resource can be
freely duplicated across multiple function applications without
entangling their results.

\paragraph{Duplicable resource and persistent modality} In other
BI-logics that support the persistent modality~\cite{Iris}, its
definition is typically based on a \emph{duplicable} resource with
respect to the product operation of the underlying BI-algebra. A
duplicable resource is idempotent with respect to the product
operation, i.e., $r \times r = r$, which is what allows us to
duplicate generator types in the type context.  In our setting,
$r \times r = r$ only holds when $r$ is a \emph{singleton} capability
containing exactly one environment. Following this idea, we define the
persistent modality to hold on formulas that do not depend on a
capability providing different choices:

\begin{definition}[Persistent Modality] We introduce a persistence
  modality $\chpers P$ to our context logic such that:
  \begin{align*}
    r \models \chpers P \text{ iff } \exists \sigma. \{\sigma\} \sqsubseteq r \land \{\sigma\} \models P
  \end{align*}\noindent
  Note that we do not require $r$ to be singleton over its entire
  variable domain, as doing so would violate Kripke monotonicity: if a
  capability $r$ is singleton, then for $r'$ such that
  $r \sqsubseteq r'$, $r'$ is not always singleton. Instead, we impose
  the more relaxed requirement that there is some
  $\{\sigma\} \sqsubseteq r$, which need only include the variables
  relevant to $P$, i.e. $\projectTo{r}{\Code{FV}(P)}$.
\end{definition}

\paragraph{Persistent type} The persistent type
$\Gamma \vdash e : \ctpers \tau$ indicates that the typing of $e$ does
not rely on the capabilities provided by either the bindings in the
type context $\Gamma$ or \emph{the reduction of the term $e$}. To
illustrate, consider the following three functions:
\begin{align*}
  &z{:}\urt{\Nat}{\top} \vdash \zzlam{x}{z} : \ort{\Nat}{\top}
    \ctsarr \urt{\Nat}{\top} \judgok \\
  &\emptyset \vdash \zzlam{x}{\Code{gen_{nat} \;()}} : \ort{\Unit}{\top}
    \ctsarr \urt{\Nat}{\top}  \judgok \\
  &\emptyset \vdash \zzlam{x}{0} \oplus
    \zzlam{x}{\Code{gen_{nat} \;() + 1}} : \ort{\Unit}{\top} \ctsarr
    \urt{\Nat}{\top}  \judgok
\end{align*}\noindent
Of these three terms, only the second can also be assigned a
persistent modality. The first is obviously not persistent since it
depends on the binding $z$ in the type context, while the reachability
properties of the third \emph{depend on how it is reduced}. To see
how, consider the following type judgment, which is semantically
invalid under \langNDname{}'s call-by-value reduction strategy:
\begin{align*}
  \emptyset \vdash \zlet{f}{\zzlam{x}{0} \oplus
  \zzlam{x}{\Code{gen_{nat} \;() + 1}}}{\zif{f\;() = 0}{f\;()}{0}} : \urt{\Nat}{\top} \quad \judgfail
\end{align*}\noindent
This term can only reach the value $0$: since $f$ is bound to either
$\zzlam{x}{0}$ or $\zzlam{x}{\Code{gen_{nat} \;() + 1}}$ when the
first let expression is reduced, two applications of $f\;()$ will
either both reduce to $0$ or some positive number. In other words, the
two applications of $f$ are indeed dependent, and $f$ is not
persistent.

\begin{definition}[Persistent Type Denotation]
  The denotation of $\ctpers \tau$ is:
  \begin{align*}
    \denot{\ctpers \tau}_\Sigma\;e \doteq \forall x.\denot{\mstep{e}{x}} \chimpl \chpers (\denot{\tau}_{\Sigma,x{:}\eraserf{\tau}}\;x) \quad (\text{fresh $x$})
  \end{align*}\noindent
  Note that if we had instead defined the type denotation as
  $\chpers (\denot{\tau}_{\Sigma}\;e)$, it would (incorrectly) contain
  inhabitants like
  $\zzlam{x}{0} \oplus \zzlam{x}{\Code{gen_{nat} \;() + 1}}$. To
  ensure that $e$ is single-valued, we introduce a universally
  quantified fresh variable $x$ that ranges over all possible
  reduction results of $e$; the persistent modality then ensures both
  the context and this $x$ are single-valued.
\end{definition}

\paragraph{Typing rules for $\ctpers$} The typing rules for persistent
types are:
\begin{align*}
\mprooftr{20mm}
      {$\Gamma \vdash e : \ctpers \tau$}
      {T-PersE}
      {$\Gamma \vdash e : \tau$}
\quad
  \mprooftr{40mm}
      {$\Gamma \vdash v : \tau$ \quad $\denot{\Gamma} \models \chpers \denot{\Gamma}$}
      {T-PersI}
      {$\Gamma \vdash v : \ctpers \tau$}
\end{align*}\noindent
The elimination rule for persistent types \textsc{T-PersE} is
derivable from \textsc{T-Sub} and the fact
$\Gamma \vdash \ctpers \tau <: \tau$ always holds.  Its introduction
rule \textsc{T-PersI} allows us to conclude that a value is a
persistent term of type $\ctpers \tau$ if we can establish that it has
type $\tau$ using a type context comprised only of persistent facts,
i.e., $\denot{\Gamma} \models \chpers \denot{\Gamma}$.

\subsection{Type Denotation for Nondeterminism} %
\label{subsec:nondet-denotation}

When reasoning about reachability properties of \langname{} terms, all
the choices available to us came from angelic refinements in the
typing context. In contrast, nondeterministic \langNDname{} terms can
themselves introduce choices that are not fully captured by our
previous type denotation. This prevents us from applying semantic
subtyping to verify properties of terms which should intuitively hold.
To illustrate, consider the following typing derivation: \smallskip
\begin{center}
\small
\begin{prooftree}
  \hypo{\emptyset \vdash \Code{gen_{bool} \;()}  : \urt{\Bool}{\top}}
  \hypo{\emptyset \vdash \urt{\Bool}{\top} \subty \ort{\Bool}{\vnu = \true} \ctoplus \ort{\Bool}{\vnu = \false}}
  \infer2[T-CtxSub]{\emptyset \vdash \Code{gen_{bool} \;()} : \ort{\Bool}{\vnu =
      \true} \ctoplus \ort{\Bool}{\vnu = \false}}
\end{prooftree}
\end{center}\smallskip

Intuitively, the derivation should be valid: $\Code{gen_{bool} \;()}$
exhibits two control flow paths, reaching \Code{true} on one and
\Code{false} on the other. Unfortunately, this conclusion does not
hold \emph{semantically} under our original denotation, i.e.,
$\not \models~ \denot{\ort{\Bool}{\vnu = \true} \ctoplus
  \ort{\Bool}{\vnu = \false}}~ \Code{gen_{bool} \;()} $. The issue
here is that the original denotation of $\choplus$ only splits the
capabilities of the \emph{context} between the two branches, not the
term itself --- for our target claim to hold, $\Code{gen_{bool} \;()}$
could only evaluate to \Code{true} under some partitioning of the
empty context, which clearly never happens!

A similar issue arises with the original denotation of function
types. The fix to both issues is to ``lift'' the extra capability
provided by a nondeterministic term into the capability context:
\begin{align*}
  \denot{x{:}\tau_x \ctsarr \tau}_\Sigma\;e
    &\doteq \forall f.\; \denot{\mstep{e}{f}}_\Sigma \chimpl
             \denot{\tau_x}_\Sigma\;x \chimpl
             \denot{\tau}_{\Sigma,f{:}\eraserf{\tau_x \ctsarr \tau}, x{:}\eraserf{\tau_x}}\;(f\;x)
  \\\denot{x{:}\tau_x \ctwand \tau}_\Sigma\;e
    &\doteq \forall f.\; \denot{\mstep{e}{f}}_\Sigma \chimpl
             \denot{\tau_x}_\Sigma\;x \chwand
             \denot{\tau}_{\Sigma,f{:}\eraserf{\tau_x\ctwand \tau}, x{:}\eraserf{\tau_x}}\;(f\;x)
  \\\denot{\tau_1 \ctoplus \tau_2}_\Sigma\;e
    &\doteq \forall x.\; \denot{\mstep{e}{x}}_\Sigma \chimpl
             \denot{\tau_1}_{\Sigma,x{:}\eraserf{\tau_1}}\;x \choplus \denot{\tau_2}_{\Sigma,x{:}\eraserf{\tau_2}}\;x
\end{align*}
In each case, a universally quantified fresh variable ($f$ or $x$)
ranges over all possible reduction results of $e$, and the original
denotation is applied to that variable.  For deterministic terms,
which have a unique reduction result, the quantifier is trivially
eliminated, so the generalized denotation coincides with the original
when $e$ is deterministic.

This new denotation allows us to lift $\Code{gen_{bool} \;()}$'s
nondeterminism into the context via the fresh variable $x$, allowing
the sum operator to partition it across the two branches as before:
\begin{align*}
  &\denot{\ort{\Bool}{\vnu = \true} \ctoplus
          \ort{\Bool}{\vnu = \false}}\;\Code{gen_{bool} \;()}
  \\=\;& \forall x.\; \Code{Atom}(x = \true \lor x = \false) \chimpl
          \demonic\,\Code{Atom}(x = \true) \choplus
          \demonic\,\Code{Atom}(x = \false)
  \quad\judgok
\end{align*}
This richer denotation also allows us to discharge the semantic
subtyping assumption in the previous typing derivation, allowing us to
verify that $\Code{gen_{bool} \;()}$ meets this intuitive
specification. Finally, we can show that \langNDname{}'s extended type
system is sound with respect to this generalized type denotation.

\begin{theorem}[\langNDname{} Type Soundness]%
  \label{thm:nondet-soundness}
  For a well-typed \langNDname{} term $\emptyset \vdash e : \tau$ and
  fresh variable $x$,
  $\{[x \mapsto v] \mid \mstep{e}{v}\} \models \denot{\tau}\;x$.
\end{theorem}

\section{Case Studies} %
\label{sec:case}

This section illustrates the expressiveness of our bunched type system
via three real-world case studies.

\paragraph{May/Must Analysis} %
\label{subsec:may-must}
\begin{figure}
\begin{subfigure}[t]{0.48\linewidth}
    \centering
\begin{minted}[xleftmargin=5pt, numbersep=4pt, linenos = true,
    fontsize = \footnotesize, escapeinside=??]{OCaml}
  val g : ?$\Code{i}{:}\ort{\Int}{\top}\ctsarr\ort{\Int}{\vnu \neq 0}$?

?$\vdash$? let f (x: int) (y: int) =
    if y >= 10 then (g x) + 1
    else x + 1
    : ?$\Code{x}{:}\ \urt{\Int}{\vnu = 0} \ctsarr \Code{y}{:}\urt{\Int}{0 \leq \vnu \leq 10} \ctwand$?
        ?$\urt{\Int}{\vnu = 1}$?
\end{minted}
    \subcaption{May-analysis helps must-analysis.}
    \label{fig:may-must-reach2}
\end{subfigure}
\hfill
\begin{subfigure}[t]{0.48\linewidth}
    \centering
\begin{minted}[xleftmargin=5pt, numbersep=4pt, linenos = true,
    fontsize = \footnotesize, escapeinside=??]{OCaml}
  val g : ?$\Code{i}{:}\ort{\Int}{\top}\ctsarr\ort{\Int}{\vnu > 0}\interty$?
          ?$\Code{i}{:}\urt{\Int}{\top}\ctsarr\urt{\Int}{\vnu > 0}$?

?$\vdash$? let f (x: int) =
    let a = g x in
    if a > 3 then 5 else 2
: ?$\Code{x}{:}\urt{\Int}{\top}\ctsarr\ort{\Int}{\vnu = 2} \ctoplus \ort{\Int}{\vnu = 5}$?
\end{minted}
    \subcaption{Must-analysis helps may-analysis.}
    \label{fig:may-must-inter}
\end{subfigure}
\vspace*{-.1in}
\caption{Compositional may- and must-style reasoning with context
  types.}
\label{fig:may-must-analysis}
\vspace*{-.1in}
\end{figure}
Prior work has considered program analyses that reason about may- and
must-properties within the same framework~\cite{BVR+08,GN+10}. Like
our type system, these analyses are compositional because callee
summaries inform caller deductions. Figure~\ref{fig:may-must-analysis}
adapts examples from \citet{GN+10}, restated in our
core language with context types assigned to both the caller
$\Code{f}$ and the callee $\Code{g}$. In our encoding, demonic types
correspond to may-analysis summaries (safety), while angelic types
correspond to must-analysis summaries (reachability).

Figure~\ref{fig:may-must-analysis}~(a) illustrates how a may-analysis
result on $\Code{g}$ enables a must-analysis result on
$\Code{f}$. \cbnewadding{$\Code{f}$'s target type asserts that it can
  return $\Code{1}$ when $\Code{x}$ can be $\Code{0}$ and $\Code{y}$
  can be any value between $\Code{0}$ and $\Code{10}$.  The
  independent arrow ensures that each input can be chosen freely from
  the other, i.e., $\Code{x}\mapsto \Code{0}$ can be paired with every
  $\Code{y}\in \Code{[0,10}]$.  The context used to type the body of
  $\Code{f}$ can thus be divided into the following partitions, each
  corresponding to one of the control flows of the function: %
  {\small
    \begin{align*}
      \Code{x}{:}\urt{\Int}{\vnu = 0}\ctstar\Code{y}{:}\urt{\Int}{0
        \leq \vnu \leq 10} \ctsubseteq{\{\Code{x,y}\}}
      &\Code{x}{:}\urt{\Int}{\vnu =
        0}\ctstar\Code{y}{:}\urt{\Int}{\vnu = 10} \interty
      \ort{\Int}{\vnu \geq 10} \\ & \ctoplus
      \Code{x}{:}\urt{\Int}{\vnu = 0}\ctstar\Code{y}{:}\urt{\Int}{0
        \leq \vnu < 10} \interty \ort{\Int}{\vnu < 10}
    \end{align*}
  }
  \noindent
  Since $x$ is independent of $y$, its angelic guarantee is preserved
  in both contexts.  Observe that the $\Code{then}$ branch can only
  return $\Code{1}$ when $\Code{g}\;\Code{x}=\Code{0}$, a possibility
  precluded by $\Code{g}$'s demonic result type. This may-analysis
  summary therefore allows us to rule out the $\Code{then}$ branch as
  a source of witnesses.  This in turn helps us focus our efforts on
  the $\Code{else}$ branch, which is able to witness that $\Code{f}$ can
  return $\Code{1}$.}

Figure~\ref{fig:may-must-analysis}~(b) illustrates how a must-analysis
result on $\Code{g}$ enables a \emph{tight} may-analysis result on
$\Code{f}$ whose type is a safety specification asserting that it
always returns either 2 or 5.  The true branch ($\Code{a} > 3$) always
returns 5, and the false branch ($\Code{a} \leq 3$) always returns 2.
Function \Code{g}'s safety specification ($\ort{\Int}{\vnu > 0}$)
establishes that $\Code{a} > 0$, ensuring neither branch produces an
out-of-range value. However, with only this type, a may analysis
cannot assert that both 2 and 5 are \emph{definitively} part of
$\Code{f}$'s behavior.  With $\Code{g}$'s must type ($\urt{\Int}{\vnu
  > 0}$), however, such an analysis can conclude that $\Code{a}$ can
reach values that are both $\leq 3$ and $>3$, guaranteeing that both
branches are definitely reachable. This makes $\Code{f}$'s result type
provably tight: the must information certifies that neither 2 nor 5 is
spurious.

\paragraph{Coverage Completeness}%
\label{subsec:coverage}
\begin{wrapfigure}{r}{.55\textwidth}
  \vspace*{-.2in}
\begin{minted}[xleftmargin=15pt, numbersep=4pt, linenos = true, fontsize = \footnotesize, escapeinside=??]{OCaml}
let rec bst_gen (r: nat) (lo: nat)
              (range: nat -> nat -> nat): nat tree =
if r <= 0 then Leaf
else Leaf ?$\oplus$?
  let r' = range (1, r) in
  let left  = bst_gen (r' - 1) lo range in
  let right = bst_gen (r - r') (lo + r') range in
  Node (lo + r') left right
\end{minted}
\vspace*{-.1in}
\caption{A BST generator adapted from~\cite{CoverageType}.}
  \label{fig:bst}
\vspace*{-.1in}
\end{wrapfigure}
The coverage type system of~\citet{CoverageType} verifies that a test
generator produces every desirable input.  Given the BST generator
shown in Figure~\ref{fig:bst}, the following specification asserts
that for any valid range generator, \texttt{bst\_gen} produces
\emph{every} BST whose keys lie within
$(\Code{lo}, \Code{lo}+r]$:\footnote{In the type for
  $\Code{bst\_gen}$, $\Code{mem}(\vnu,x)$ is a \emph{method
    predicate}, i.e., an uninterpreted function used to encode
  properties of algebraic data types like trees, in this case a tree
  membership property.}
\begin{align*}
  \mykw{val}\;\Code{bst\_gen} : \ctpers(&
    \Code{r}{:}\ort{\Nat}{\top} \ctsarr \Code{lo}{:}\ort{\Nat}{\top} \ctsarr
  \\&
    \Code{range}{:}\ctpers (x{:}\ort{\Nat}{\top} \ctsarr
      y{:}\ort{\Nat}{x < \vnu} \ctsarr
      \urt{\Nat}{x < \vnu < y})
    \ctsarr
  \\ & \quad
    \urt{\Nat\;\Code{tree}}{\Code{bst}(\vnu) \land
      \forall x.\, \Code{mem}(\vnu, x) \iff
      (\Code{lo} < x \leq \Code{lo}+r)})
\end{align*}
In our setting, input generators always introduce fresh angelic
capabilities at each invocation, independently of the surrounding
context, for \emph{arbitrary} input parameters. Input generators thus
have context types of the following shape:
\begin{align*}
    \overbrace{\ctpers (\quad \underbrace{\overline{x{:}\ort{b_x}{\phi_x}\ctsarr}}_{\texttt{demonic parameters}} \quad \underbrace{\urt{b}{\phi}}_{\texttt{angelic result}} \quad )}^{\texttt{duplicatable}}
\end{align*}\noindent
The \Code{range} parameter of \Code{bst\_gen}, for example, is
generator is persistent and has demonic parameter types
($\ort{\Nat}{\top}$ and $\ort{\Nat}{x < \vnu}$) and an angelic result
type ($\urt{\Nat}{x < \vnu < y}$).  The demonic parameters allow
successive calls to \Code{range} with different arguments to be typed
with independent result capabilities --- the generator's output at
each recursive call is disjoint from its outputs at other calls.  The
BST generator follows the same pattern: its demonic parameter types
allow any valid size $\Code{r}$ and lower bound $\Code{lo}$, and its
angelic result type asserts that every valid BST is reachable by some
sequence of generator choices.

\paragraph{Incorrectness Reasoning} %
\label{subsec:disprove}
\begin{figure}[h!]
  \centering
\vspace*{-.15in}
\begin{minted}[xleftmargin=5pt, numbersep=4pt, linenos = true, fontsize = \footnotesize, escapeinside=??]{OCaml}
    let rec merge (l1: nat list) (l2: nat list) : nat list =
        match l1, l2 with
        | l1, [] -> l1
        | [], l2 -> l2
        | h1 :: t1, h2 :: t2 ->
            let t = merge t1 t2 in (* t contains elements in t1 *)
            if h1 < h2 then h1 :: h2 :: t (* h2 may be greater than elements in t1 *)
            else if h1 > h2 then h2 :: h1 :: t
            else h1 :: t
\end{minted}
\vspace*{-.1in}
\caption{A function that merges two sorted lists. If either of
the input lists are empty, \Code{merge} returns the other (lines
$3-4$). Otherwise, the function merges the tails of the two inputs and appends the two top elements ahead of the merged result. }
\label{fig:merge}
\vspace*{-.1in}
\end{figure}
When a refinement type system fails to prove a safety property, the
target program is not necessarily incorrect: may-style
(overapproximate) reasoning offers no guarantee that a presumed
erroneous behavior is actually reachable. Recent
work~\cite{RefinementTypeRefutations, RefinementTypeSymbolicExecution}
addresses this limitation by using symbolic execution to search for
counterexamples; our context type system can establish such
counterexamples directly.  Consider the $\Code{merge}$ function shown
in Figure~\ref{fig:merge}. When used in a mergesort procedure, this
function is intended to merge two sorted lists and produce a sorted
output list:
\begin{align*}
  \vdash \Code{merge}~:{}&\Code{l1}{:}\ort{\List[\Code{nat}]}{\I{sorted}(\vnu)}
  \ctsarr \Code{l2}{:}\ort{\List[\Code{nat}]}{\I{sorted}(\vnu)} \\
  &\qquad \ctsarr \ort{\List[\Code{nat}]}{\I{sorted}(\vnu)} \quad \judgfail
\end{align*}\noindent
Verification fails nonetheless. The base cases (lines $3$--$4$)
type-check, so the failure arises in the recursive case, but the
refinement type system offers little guidance on the underlying fault,
since it cannot identify which execution branch leads to the unsorted
output.  A symbolic execution engine might identify the following
concrete counterexample:
\begin{align*}
    \texttt{Counterexample:}&\quad  \text{ when } \Code{l1} = [1;2] \text{ and } \Code{l2} = [3;4] \text{ then } \mstep{\Code{merge}\;[1;2]\;[3;4]}{[1;3;2;4]}
\end{align*}\noindent
The failure stems from the assumption that, after comparing
$\Code{h1}$ and $\Code{h2}$, $\Code{h2}$ is always the second-smallest
element (line $7$).  Even when $\Code{h1} < \Code{h2}$, however,
$\Code{h2}$ may still be larger than the next element of $\Code{l1}$,
breaking sortedness.  Rather than appealing to a separate symbolic
execution procedure to discover this counterexample, we can directly
prove that \Code{merge} operates incorrectly, by type-checking it with
the following type:
\begin{align*}
  &\Code{l1}{:}\urt{\List[\Code{nat}]}{\I{sorted}(\vnu)}
  \ctwand \Code{l2}{:}\urt{\List[\Code{nat}]}{\I{sorted}(\vnu)} \\
  &\qquad \ctwand \ort{\List[\Code{nat}]}{\neg \I{sorted}(\vnu)}
  \ctoplus \ort{\List[\Code{nat}]}{\top}
\end{align*}\noindent
This type follows the ``affine reachability'' style of
Section~\ref{sec:overview}: the result type is a sum of the top type
$\ort{\List[\Code{nat}]}{\top}$ and the negation of the original
safety property, $\ort{\List[\Code{nat}]}{\neg \I{sorted}(\vnu)}$,
that captures the erroneous behavior.  The parameter types are given
\emph{disjoint angelic types}, indicating that $\Code{l1}$ and
$\Code{l2}$ may independently range over arbitrary sorted lists.
Without this disjointness, a carefully crafted context can hide the
bug, as shown below. Here $\Code{l1}$ and $\Code{l2}$ each
\begin{wrapfigure}{l}{.4\textwidth}
  \begin{center}
    \vspace*{-.17in}
\begin{minted}[xleftmargin=5pt, numbersep=4pt, linenos = true, fontsize = \footnotesize, escapeinside=??]{OCaml}
  let l1 = sorted_list_gen () in
  let l2 = match l1 with
           | [] -> sorted_list_gen ()
           | _ -> [] in
  merge l1 l2
\end{minted}
\end{center}
\vspace*{-.1in}
\end{wrapfigure}
range over all sorted lists, given by $\Code{sorted\_list\_gen}$, yet
they cannot both be non-empty simultaneously.  Under this environment,
the bug in \Code{merge} cannot be triggered, since the case where both inputs
are non-empty (line $5$) is never reached.

\cbnewadding{To establish that \Code{merge}
can produce an unsorted result, we first typecheck it against the following type:
\begin{align*}
  &\Code{l1}{:}\ort{\List[\Code{nat}]}{\I{sorted}(\vnu)}
  \ctsarr \Code{l2}{:}\ort{\List[\Code{nat}]}{\I{sorted}(\vnu)} \\
  &\qquad \ctwand \textcolor{magenta}{\mbox{\rm\textbf{\{}}} \vnu{:}\;\List[\Code{nat}] \;|\; (l_1 = [1;2] \land l_2 = [3;4] \impl \vnu = [1;3;2;4]) \textcolor{magenta}{\mbox{\rm\textbf{\}}}}
  \\&\qquad\qquad \ctoplus \textcolor{magenta}{\mbox{\rm\textbf{\{}}} \vnu{:}\;\List[\Code{nat}] \;|\; (l_1 = [] \land l_2 = [] \impl \vnu = []) \land (l_1 = [2] \land l_2 = [4] \impl \vnu = [2;4]) \textcolor{magenta}{\mbox{\rm\textbf{\}}}}
\end{align*}\noindent
Its demonic parameters allow us to establish the function's behavior
for arbitrary sorted inputs, facilitating the use of \textsc{T-Fix}
and the typing of recursive calls within the function body.
The left result summand captures the counterexample above and
establishes the target unsorted refinement when the context selects
those inputs. The right result summand records the input-output
relationships needed for the recursive calls and can be weakened to
$\ort{\List[\Code{nat}]}{\top}$. Semantic subsumption then yields
the original type.
}

\section{Discussion}%
\label{sec:discussion}

\cbnewadding{The motivation for our type system is primarily
  foundational: our goal is to provide a unified framework for safety
  and reachability reasoning in a functional language. Consequently,
  we have prioritized issues related to semantic expressivity over
  more practical considerations. By doing so, our type system allows
  demonic and angelic modalities to be freely combined within the same
  type, context, and typing judgement, relying on bunched contexts to
  record whether the choices associated with different variables are
  entangled or independent.  }

Our context type system combines the strengths of refinement type
systems and BI-logic-based verification approaches, enabling
compositional reasoning about semantic resources (in our case, angelic
and demonic capabilities) from expressive, modular specifications of
program behavior. It also inherits many of the challenges of these
systems, including the need for some form of user-provided guidance
and the inherent tension between richness of specification and
decidability.

For example, a syntax-directed typing algorithm, necessary for any
efficient implementation, would require additional typing annotations
to constrain the flexibility of the existing typing rules.
\cbnewadding{Several aspects of the declarative system presented here
  complicate such an algorithm. First, the syntax of a term does not
  determine the bunched structure of its context: let expressions, for
  example, can be typed using either the \textsc{T-Let} or
  \textsc{T-LetI} rules, and the choice dictates whether the contexts
  are entangled or independent. Similarly, rules such as
  \textsc{T-Match} require constructing appropriate context partitions
  for each reachable branch. Second, the modality of a target result
  type does not by itself determine the modalities against which
  subterms should be checked: the parameter type of an applied
  function, for example, may require its argument to be checked
  angelically even when the result is checked against a demonic type.}
It is impractical, however, to ask users to write function signatures
that anticipate the requirements imposed by all calling
contexts. \cbnewadding{Even putting aside the question of typing
  annotation burden, the \textsc{T-Sub} and \textsc{T-CtxSub} rules
  delegate substantial reasoning to semantic subsumption relations;
  any type-checking procedure must be able to automatically discharge
  the obligations that arise from these rules.} Allowing unrestricted
angelic and demonic modalities in the typing context can furthermore
lead to formulae with quantifier alternations that lie outside
decidable fragments of first-order logic.

We note that similar challenges arise in separation logic-based
verification tools, with initial proposals requiring substantial
user-supplied annotations for tasks such as frame inference and
invariant discovery. Subsequent work has shown that these costs can be
substantially reduced through automated inference techniques, e.g.,
bi-abduction~\cite{CD11}, which enables scalable precondition and
frame inference, and that underpin modern automated separation logic
analyzers~\cite{DFLO19}.  \cbnewadding{The case studies of
  \autoref{sec:case} illustrate potential applications of the general
  foundation developed in this work.  Moreover, restricting where
  angelic and demonic modalities can appear has previously been shown
  to yield effective verification procedures: for example, Coverage
  Types~\cite{CoverageType} provide a restricted instance in which
  function parameters must be demonic.  Identifying other useful
  restrictions and investigating typing algorithms for context types
  in their full generality are important directions for future work.}
Alternatively, rather than pursuing full automation, we might consider
providing support for context types within an interactive mechanized
verification system like Rocq in the spirit of frameworks like
Iris~\cite{Iris} or VST~\cite{AD+14}.  In this setting, users would
retain fine-grained control over typing choices, with specialized
context type-aware tactics used to facilitate local type inference,
search, and backtracking.

\section{Related Work}%
\label{sec:related}

\paragraph{Bunched typing and bunched implication.}

The idea of expressing fine-grained resource control within a program
logic as introduced in Bunched Implication~\cite{OHP99} is central to
modern-day program verifiers~\cite{AD+14,Iris,SLK+21,MSJ16,GM+25}
based on separation logic~\cite{SeparationLogic,OH19}.  Transplanting
the notion of \emph{bunched typing} as introduced
by~\citet{BunchedType}, using $\ctcomma$ and $\ctstar$ to track
entangled and independent resource use within a type context has
received much less attention.  Recently proposed type systems and
type-based verification frameworks that support resource
reasoning~\cite{FR+21,PLM+23,JJKD18} do so by exposing an explicit
ownership discipline, rather than manipulating a bunched context
structure.  Notably, these works are designed primarily with safety in
mind, and do not consider the role of resources and contexts as a
unifying mechanism for safety and reachability.

\paragraph{Reachability verification.}
Recent work on reachability verification in type systems includes
two-sided type systems~\cite{Ill-types} which refute typing formulas
to show that ill-formed programs do not evaluate, and a complementary
approach~\cite{RamsayIncorType} that internalizes negation directly as
a complement operator on types to support incorrectness reasoning.
Coverage types~\cite{CoverageType} underapproximate
test-input-generator behavior to establish coverage completeness.
\cbnewadding{\citet{UnnoST18} also combine universal and existential
  reasoning in a refinement type system for nondeterministic
  higher-order programs.  Their existential type $\{\vnu{:}b \mid
  \phi\}^{\exists\exists}$ states that some execution produces a value
  satisfying $\phi$, whereas our angelic refinement $\urt{b}{\phi}$
  guarantees that every value satisfying $\phi$ is reachable, a
  strictly stronger guarantee.  Moreover, their flat typing
  environments do not record whether existentially quantified bindings
  are entangled or independent.  For example, existential bindings for
  both $x$ and $y$ cannot distinguish the contexts
  $\Gamma_{\mathit{en}}$ and $\Gamma_{\mathit{dis}}$ in
  \autoref{eqn:Ctxen} and \autoref{eqn:Ctxdis}: they assert the
  existence of suitable witnesses, but not that all independent
  combinations of those witnesses are available.  Consequently, their
  system has no counterpart of the independent arrow in the type of
  \ref{ex:div}, which requires the choices for its second argument to
  be independent of those for its first.}  To summarize, while these
prior efforts emphasize an algorithmic characterization of
reachability within a type system, they do not consider the logical
underpinning that relates reachability and safety claims as we do
here, and instead treat reachability properties independently of other
safety considerations.

We note that contemporary work on \emph{Reachability
  Types}~\cite{ReachabilityTypes, WBJBR+24} uses ``reachability'' in a
sense unrelated to how it is used here, and focuses on soundly
tracking aliasing of mutable references in higher-order effectful
programs.

\paragraph{Incorrectness logics.}
Our angelic modality can be viewed as establishing a witness to a
property, and thus is closely related to operators found in
underapproximate logics~\cite{ReverseHoareLogic, OHP19, RBD+20,
  RHLE, HyperHoareLogic, QuantitativeWeakestHyperPre, UTurn,
  BackwardIncorrectness} that establish judgements which guarantee the
existence of certain executions rather than ensuring the absence of
unsafe executions.

However, these logics and their descendants either fix a single
approximation direction per triple~\cite{ReverseHoareLogic, OHP19,
  RBD+20} or disallow approximation entirely~\cite{ESL+23}, and do not
support the construction or decomposition of context structure, which
is a central feature of our type system. A complementary line of work
has studied the relationship between various correctness and
incorrectness program logics, when viewed through the lens of
predicate transformers and Galois connections~\cite{Cousot+24, VK+25};
these efforts relate distinct program logics to one another
externally, whereas our type system internalizes both approximation
directions into a single reasoning framework.

Like us, outcome logic~\cite{ZDS23, Zil25TOPLAS} also unifies safety
and reachability through branch composition $\varphi \oplus \psi$, and
its $\oplus$ adjoins control-flow alternatives like our $\ctoplus$.
But, mixing angelic and demonic constraints within a single outcome
logic judgement requires an \emph{a priori} partitioning of outcomes
via $\oplus$, whereas our type system determines valid angelic choices
compositionally, without requiring such a partition to be specified
upfront. Outcome logic has since been extended to support
separation logic-style local reasoning~\cite{ZSS+24} and programs
featuring adversarially resolved nondeterminism~\cite{ZKST+25}. In
these extensions, however, ``demonic'' and ``angelic'' characterize
how a program's nondeterminism is resolved by its semantics, whereas
our modalities are specification-level annotations that determine
whether an individual refinement is read as an upper or lower bound,
and may be mixed freely within a single type judgement.

\paragraph{May- and must-style program analysis.}
There has been significant prior work on combining may- and must-style
reasoning within a program analysis~\cite{GHK+06,GN+10,NYC+12,BVR+08}.
For example,~\citet{GN+10} alternates overapproximate may-summaries
with underapproximate must-summaries so that callee information can
inform caller analysis.

Refutation is a type-level specialization of may/must reasoning: a
failed refinement type check (which is may-style) does not by itself
establish that a bug is reachable (which requires must-style
evidence).  Refinement type
refutations~\cite{RefinementTypeRefutations} address this gap by
deriving \emph{must}-style witnesses that explain why typing failed,
often through concrete counterexample instantiations, while
symbolic-execution tooling~\cite{RefinementTypeSymbolicExecution}
searches for such counterexamples externally.  More broadly, our
context types provide a unified theoretical foundation for both
approaches by exploiting the duality between safety and reachability
reasoning.

\paragraph{Context-aware typing}
Coeffect and graded modal type systems~\cite{Coeffects, Katsumata+14,
  GaboardiCombining, OrchardQuantitative} annotate programs with
context information such as usage, liveness, and security levels,
interpreted via graded (co)monads.  Like our context types, they treat
contextual information as a typed resource with a compositional
algebra, but they track the demands a computation makes on its
environment or what effects the computation has on its environment,
rather than how a typing context \emph{interprets} value sets under
demonic or angelic modalities.  Thus, they do not provide a mechanism
to internalize safety/reachability modalities on the same base
refinement, nor do they provide a bunched capability structure for
tracking dependencies between bindings in a higher-order setting.

\section{Conclusion} %
\label{sec:conclusion}

This paper presented context types, a new kind of type refinement in
which every type qualifier is equipped with a modality indicating
whether it should be interpreted as an over- or under-approximate
specification. The resulting type system permits both kinds of
specifications to be compositionally combined within a unified
framework for safety and reachability verification. The key insight is
to treat a typing context as a contextual capability --- a resource
whose structure governs the demonic and angelic choices available when
reasoning about terms. Building on this insight, we have developed a
resource algebra of capabilities and a modal context logic over that
algebra, and given a denotation of our type system into context logic.
Our metatheory, mechanized in Rocq, establishes that a single
soundness theorem delivers both guarantees: every value produced by a
well-typed term satisfies its demonic types, and every value
prescribed by its angelic types is witnessed by some execution.

\section*{Disclosure of Use of AI in Research}
All definitions and theorem statements in the accompanying Rocq
artifact were written by the authors. The authors used OpenAI Codex to
help construct and debug portions of the corresponding proof scripts
of those theorems, e.g., asking it to suggest proof steps and propose
edits to proof scripts. All AI-generated suggestions were reviewed by
the authors and revised as needed, and the final proof scripts were
checked by the Rocq kernel. The authors are responsible for the
correctness of the definitions, theorem statements, proofs, and
artifact.

\section*{Data Availability Statement}
The complete Rocq formalization and mechanized proofs accompanying this
paper are available at \url{https://zenodo.org/records/23293018}.

\bibliographystyle{ACM-Reference-Format}
\bibliography{bibliography}

\ifdefined\paperonly
\else
\newpage
\appendix
\section{Overview of Supplemental Materials}

The supplemental material is organized as follows.
\autoref{sec:tech:semantics} gives the complete operational semantics
of \langname{}.  \autoref{sec:tech:basic-typing} provides the full set
of \langname{}'s basic typing rules, qualifier typing rules, and
well-formedness judgments for context types and type contexts.
\autoref{sec:formalization-details} provides the details definition of
environment and capability.  A Rocq formalization of our
nondeterministic extensions to the core language, \langNDname{}, and
its type system, together with machine-checked proofs of the
fundamental theorem and soundness theorem, is available at
\url{https://anonymous.4open.science/r/POPL27-ContextType-SM-52EC}.

\section{Operational Semantics}\label{sec:tech:semantics}

\begin{figure}[h!]
 {\small
  {\normalsize
\begin{flalign*}
 &\text{\textbf{Operational Semantics }} & \fbox{$e \hookrightarrow e$}
\end{flalign*}
}
\\ \
\begin{prooftree}
\hypo{op\ \overline{v} \Downarrow v_y }
\infer1[\textsc{StAppOp}]{
\zlet{y}{op\ \overline{v}}{e} \hookrightarrow e[y\mapsto v_y]
}
\end{prooftree}
\\ \ \\ \ \\
\begin{prooftree}
\hypo{e_1 \hookrightarrow e_1'}
\infer1[\textsc{StLetE1}]{
\zlet{y}{e_1}{e_2} \hookrightarrow \zlet{y}{e_1'}{e_2}
}
\end{prooftree}
\quad
\begin{prooftree}
\hypo{}
\infer1[\textsc{StLetE2}]{
\zlet{y}{v}{e} \hookrightarrow e[y\mapsto v]
}
\end{prooftree}
\\ \ \\ \ \\
\begin{prooftree}
\hypo{}
\infer1[\textsc{StLetAppLam}]{
\zlet{y}{\zzlam{x}{e_1}\ v_x}{e_2} \hookrightarrow \zlet{y}{e_1[x\mapsto v_x]}{e_2}
}
\end{prooftree}
\\ \ \\ \ \\
\begin{prooftree}
\hypo{}
\infer1[\textsc{StLetAppFix}]{
\zlet{y}{\zzfix{f}{x}{e_1}\ v_x}{e_2} \hookrightarrow
\zlet{y}{(\zzlam{f}{e_1[x\mapsto v_x]}) \ (\zzfix{f}{x}{e_1})}{e_2}
}
\end{prooftree}
\\ \ \\ \ \\
\begin{prooftree}
\hypo{}
\infer1[\textsc{StMatch}]{
\match{d_i \ \overline{v_j}} \overline{d_i\ \overline{y_j} \to e_i} \hookrightarrow e_i[\overline{y_j \mapsto v_j}]
}
\end{prooftree}
}
\caption{Small Step Operational Semantics}
    \label{fig:semantics}
\end{figure}

The small-step operational semantics of \langname{} are shown in
\autoref{fig:semantics}. The rules are parameterized over the
semantics of built-in operators
$\primop\; \overline{v} \Downarrow v_y$.

\newpage
\section{Basic Typing and Well-Formedness Rules}\label{sec:tech:basic-typing}

\begin{figure}[ht!]
  {\small
  \begin{align*}
    \text{\textbf{Type erasure}} \quad & \eraserf{\ort{b}{\phi}} \doteq b \quad \eraserf{\urt{b}{\phi}} \doteq b \quad \eraserf{\tau_1 \ctinter \tau_2} \doteq \eraserf{\tau_1}
    \\&\eraserf{\tau_1 \ctoplus \tau_2} \doteq \eraserf{\tau_1} \quad  \eraserf{x{:}\tau_x \ctsarr \tau} \doteq \eraserf{\tau_x} \sarr \eraserf{\tau} \quad
    \eraserf{x{:}\tau_x \ctwand \tau} \doteq \eraserf{\tau_x} \sarr \eraserf{\tau} \quad
    \eraserf{\ctpers \tau} \doteq \eraserf{\tau}
    \\& \eraserf{\emptyset} \doteq \emptyset \quad \eraserf{x{:}\tau} \doteq x{:}\eraserf{\tau} \quad \eraserf{\Gamma_1 \ctcomma \Gamma_2} \doteq \eraserf{\Gamma_1}, \eraserf{\Gamma_2} \quad \eraserf{\Gamma_1 \ctstar \Gamma_2} \doteq \eraserf{\Gamma_1}, \eraserf{\Gamma_2}  \quad \eraserf{\Gamma_1 \ctoplus \Gamma_2} \doteq \eraserf{\Gamma_1}
 \end{align*}}
\caption{Type and Typing Context Erasure}
\label{fig:type-erasure}
\end{figure}

\begin{figure}[ht!]
    {\footnotesize
    {\small
   \begin{flalign*}
    &\text{\textbf{Basic Typing }} & \fbox{$\Sigma \basicvdash e : s$}
   \end{flalign*}}
   \begin{prooftree}
   \hypo{}
   \infer1[\textsc{BtConst}]{
   \Sigma \basicvdash c : \Code{BasicTy}(c)
   }
   \end{prooftree}
   \quad
   \begin{prooftree}
   \hypo{ \Sigma(x) = s }
   \infer1[\textsc{BtVar}]{
   \Sigma \basicvdash x : s
   }
   \end{prooftree}
   \quad
   \begin{prooftree}
   \hypo{\Sigma, x{:}s_x \basicvdash e : s}
   \infer1[\textsc{BtFun}]{
   \Sigma \basicvdash \lambda x{:}s_x.e : s_x{\sarr}s
   }
   \end{prooftree}
   \\[.8em]
   \begin{prooftree}
   \hypo{\Sigma, f{:}b\sarr s \basicvdash \lambda x{:}b.e : b\sarr s}
   \infer1[\textsc{BtFix}]{
   \Sigma \basicvdash \zfix{f}{b\sarr s}{x}{b}{e}
   : b{\sarr}s
   }
   \end{prooftree}
   \quad
   \begin{prooftree}
   \hypo{\Code{BasicTy}(\primop) = \overline{s_i}{\sarr}s \quad \forall i. \Sigma \basicvdash v_i : s_i}
   \infer1[\textsc{BtPrimOp}]{
   \Sigma \basicvdash \primop\, \overline{v_i} : s
   }
   \end{prooftree}
   \\[.8em]
   \begin{prooftree}
   \hypo{\Sigma \basicvdash v_1 : s_2 {\sarr} s \quad \Sigma \basicvdash v_2 : s_2}
   \infer1[\textsc{BtApp}]{
   \Sigma \basicvdash v_1\,v_2 : s
   }
   \end{prooftree}
   \\[.8em]
   \begin{prooftree}
   \hypo{\emptyset \basicvdash e_1 : s_x \quad \Sigma, x{:}s_x \basicvdash e_2 : s}
   \infer1[\textsc{BtLetE}]{
   \Sigma \basicvdash \zlet{x:s_x}{e_1}{e_2} : s
   }
   \end{prooftree}
   \quad
   \begin{prooftree}
   \hypo{\Sigma \basicvdash v : s_v \quad
   \forall i, \Code{BasicTy}(d_i) = \overline{s_j} {\sarr} s_v \quad
   \Sigma, \overline{y_j{:}s_j} \basicvdash e_i : s}
   \infer1[\textsc{BtMatch}]{
   \Sigma \basicvdash \match{v} \overline{d_i\,\overline{y_j} \to e_i} : s
   }
   \end{prooftree}
   }
   \caption{Basic Typing Rules}
       \label{fig:basic-type-rules}
   \end{figure}

   \begin{figure}[ht!]
    {\footnotesize
    {\small
   \begin{flalign*}
    &\text{\textbf{Basic Qualifier Typing }} & \fbox{$\Sigma \basicvdash \phi: s$}
   \end{flalign*}}
   \begin{prooftree}
   \hypo{\Code{BasicTy}(c) = s }
   \infer1[\textsc{BtConst}]{
   \Sigma \basicvdash c : s
   }
   \end{prooftree}
   \quad
   \begin{prooftree}
   \hypo{ \Sigma(x) = s }
   \infer1[\textsc{BtVar}]{
   \Sigma \basicvdash x : s
   }
   \end{prooftree}
   \quad
   \begin{prooftree}
   \hypo{ }
   \infer1[\textsc{BtTop}]{
   \Sigma \basicvdash \top : \Code{bool}
   }
   \end{prooftree}
   \quad
   \begin{prooftree}
   \hypo{ }
   \infer1[\textsc{BtBot}]{
   \Sigma \basicvdash \bot : \Code{bool}
   }
   \end{prooftree}
   \\[.8em]
   \begin{prooftree}
   \hypo{\Code{BasicTy}(\primop) = \overline{s_i}{\sarr}s \quad \forall i. \Sigma \basicvdash \phi_i : s_i }
   \infer1[\textsc{BtOp}]{
   \Sigma \basicvdash \primop\,\overline{\phi_i} : s
   }
   \end{prooftree}
   \quad
   \begin{prooftree}
   \hypo{
   \Sigma \basicvdash \phi : \Code{bool} }
   \infer1[\textsc{BtNeg}]{
   \Sigma \basicvdash \neg \phi : \Code{bool}
   }
   \end{prooftree}
   \\[.8em]
   \begin{prooftree}
   \hypo{
   \Sigma \basicvdash \phi_1 : \Code{bool} \quad
   \Sigma \basicvdash \phi_2 : \Code{bool} }
   \infer1[\textsc{BtAnd}]{
   \Sigma \basicvdash \phi_1 \land \phi_2 : \Code{bool}
   }
   \end{prooftree}
   \quad
   \begin{prooftree}
   \hypo{
   \Sigma \basicvdash \phi_1 : \Code{bool} \quad
   \Sigma \basicvdash \phi_2 : \Code{bool} }
   \infer1[\textsc{BtOr}]{
   \Sigma \basicvdash \phi_1 \lor \phi_2 : \Code{bool}
   }
   \end{prooftree}
   \quad
   \begin{prooftree}
   \hypo{
   \Sigma, x{:}b \basicvdash \phi : \Code{bool} }
   \infer1[\textsc{BtForall}]{
   \Sigma \basicvdash \forall x{:}b. \phi : \Code{bool}
   }
   \end{prooftree}
   }
   \caption{Basic Qualifier Typing Rules}
       \label{fig:basic-qualifier-type-rules}
   \end{figure}

\begin{figure}[ht!]
    {\footnotesize
    {\small
   \begin{flalign*}
    &\text{\textbf{Well-Formedness }} & \fbox{$\Sigma \wfvdash \tau$ \quad $\Sigma \wfvdash \Gamma$}
   \end{flalign*}}
   \begin{prooftree}
   \hypo{\Sigma, \vnu{:}b \basicvdash \phi : \Bool}
   \infer1[\textsc{WfBaseOver}]{
    \Sigma \wfvdash \ort{b}{\phi}
   }
   \end{prooftree}
   \quad
   \begin{prooftree}
    \hypo{\Sigma, \vnu{:}b \basicvdash \phi : \Bool}
    \infer1[\textsc{WfBaseUnder}]{
     \Sigma \wfvdash \urt{b}{\phi}
    }
    \end{prooftree}
    \\[.8em]
    \begin{prooftree}
        \hypo{\Sigma \wfvdash \tau_x \quad \Sigma, x{:}\eraserf{\tau_x} \wfvdash \tau}
        \infer1[\textsc{WfArr}]{
         \Sigma \wfvdash x{:}\tau_x \ctsarr \tau
        }
    \end{prooftree}
        \quad
    \begin{prooftree}
        \hypo{\Sigma \wfvdash \tau_x \quad \Sigma \wfvdash \tau}
        \infer1[\textsc{WfWand}]{
            \Sigma \wfvdash x{:}\tau_x \ctwand \tau
        }
    \end{prooftree}
    \quad
    \begin{prooftree}
        \hypo{\Sigma \wfvdash \tau}
        \infer1[\textsc{WfPers}]{
            \Sigma \wfvdash \ctpers \tau
        }
    \end{prooftree}
    \\[.8em]
    \begin{prooftree}
        \hypo{\Sigma \wfvdash \tau_1 \quad \Sigma \wfvdash \tau_2 \quad \eraserf{\tau_1} = \eraserf{\tau_2}}
        \infer1[\textsc{WfInter}]{
         \Sigma \wfvdash \tau_1 \interty \tau_2
        }
    \end{prooftree}
    \quad
    \begin{prooftree}
        \hypo{\Sigma \wfvdash \tau_1 \quad \Sigma \wfvdash \tau_2 \quad \eraserf{\tau_1} = \eraserf{\tau_2}}
        \infer1[\textsc{WfSum}]{
         \Sigma \wfvdash \tau_1 \ctoplus \tau_2
        }
    \end{prooftree}
    \\[.8em]
    \begin{prooftree}
        \hypo{}
        \infer1[\textsc{WfEmp}]{
            \Sigma \wfvdash \emptyset
        }
    \end{prooftree}
    \quad
        \begin{prooftree}
            \hypo{\Sigma \wfvdash \tau_x \quad x \notin \Dom{\Sigma}}
            \infer1[\textsc{WfSingle}]{
             \Sigma \wfvdash x{:}\tau_x
            }
        \end{prooftree}
    \\[.8em]
    \begin{prooftree}
        \hypo{\Sigma \wfvdash \Gamma_1 \quad \Sigma, \eraserf{\Gamma_1} \wfvdash \Gamma_2}
            \infer1[\textsc{WfComma}]{
             \Sigma \wfvdash \Gamma_1 \ctcomma \Gamma_2
        }
    \end{prooftree}
    \quad
    \begin{prooftree}
        \hypo{\Sigma \wfvdash \Gamma_1 \quad \Sigma \wfvdash \Gamma_2}
            \infer1[\textsc{WfStar}]{
             \Sigma \wfvdash \Gamma_1 \ctstar \Gamma_2
        }
    \end{prooftree}
    \\[.8em]
    \begin{prooftree}
        \hypo{\Sigma \wfvdash \Gamma_1 \quad \Sigma \wfvdash \Gamma_2 \quad \eraserf{\Gamma_1} = \eraserf{\Gamma_2}}
            \infer1[\textsc{WfCtxSum}]{
             \Sigma \wfvdash \Gamma_1 \ctoplus \Gamma_2
        }
    \end{prooftree}
   }
   \caption{Well-Formedness Rules for Context Types and Type Contexts}
       \label{fig:basic-refinement-type-rules}
   \end{figure}

   The type erasure function is defined in \autoref{fig:type-erasure}.
   The basic typing rules of our core language, qualifiers and
   refinement types are shown in \autoref{fig:basic-type-rules},
   \autoref{fig:basic-qualifier-type-rules}, and
   \autoref{fig:basic-refinement-type-rules}, resp. We use an
   auxiliary function $\Code{BasicTy}$ to provide a basic type for the
   primitives of our language, e.g., constants, built-in operators,
   and data constructors.

\newpage
\section{Semantic Domain}%
\label{sec:formalization-details}

In this section, we provide the complete definitions of environments
and capability contexts.

\begin{definition}[Environment]\label{def:env}
  Assume $\Var$ and $\Val$ range over variable and value domains,
  resp. An environment $\sigma$ is a partial map from variables to
  values, i.e., $\Var \rightharpoonup \Val$.
  We define the following standard operations on environments:
\begin{itemize}
 \item \emph{Domain}: We write $\Dom{\sigma}$ for its domain, i.e., the set of variables where the store is defined.
  \item \emph{Compatibility}: Write $\mapCompat{\sigma_1}{\sigma_2}$ when $\sigma_1$ and $\sigma_2$ are
    \emph{compatible}, i.e.\ they agree on the intersection of their domains.
    \[
      \mapCompat{\sigma_1}{\sigma_2}
      \doteq
      \forall x \in \Dom{\sigma_1} \cap \Dom{\sigma_2}.\; \sigma_1(x) = \sigma_2(x)
    \]
    Note that the concatenation of two environment is defined as a map
    union $\sigma_1 \cup \sigma_2$, which is defined when
    $\mapCompat{\sigma_1}{\sigma_2}$ holds.
  \item \emph{Projection}: Fix finite $X \subseteq_{\mathrm{fin}}
    \Var$ and an environment $\sigma$; we write $\projectTo{\sigma}{X}$ for
    the projection of $\sigma$ to $\Dom{\sigma} \cap X$, i.e.,
  \begin{align*}
    \projectTo{\sigma}{X} \doteq \{ [x\mapsto \sigma(x)] \mid x \in \Dom{\sigma} \cap X \}
  \end{align*}
  \item \emph{Refinement}: The projection operation can derive a
    partial order refinement relation. Write $\sigma_1 \sqsubseteq
    \sigma_2$ when $\sigma_1$ is a refinement of $\sigma_2$, i.e.,
    $\sigma_1 = \projectTo{\sigma_2}{\Dom{\sigma_1}} $.
\end{itemize}
\end{definition}

\begin{definition}[Contextual capability]\label{def:contextual-capability-details}
A contextual capability $r$ is a \emph{non-empty} set of environments share the same domain.
\begin{itemize}
  \item \emph{Domain}: $\Dom{r} \doteq \Dom{\sigma}$ for arbitrary $\sigma \in r$, since $r$ is non-empty and shares the same domain.
  \item \emph{Compatibility}: Write $\mapCompat{r_1}{r_2}$ when all stores in $r_1$ are compatible with all stores in $r_2$, i.e.,
    \[
      \mapCompat{r_1}{r_2}
      \doteq
      \forall \sigma_1 \in r_1.\;\forall \sigma_2 \in r_2.\; \mapCompat{\sigma_1}{\sigma_2}
    \]
  \item \emph{Projection}: The projection operation $\projectTo{r}{X}$ is defined piecewise, i.e.,
  \begin{align*}
    \projectTo{r}{X} \doteq \{ \projectTo{\sigma}{X} \mid \sigma \in r \}
  \end{align*}
  \item \emph{Refinement}: the refinement relation $r_1 \sqsubseteq r_2$ holds iff $r_1 = \projectTo{r_2}{\Dom{r_1}} $.
\end{itemize}
\end{definition}

\begin{definition}[Contextual Capabilities Product]\label{def:contextual-capabilities-product}
  The product of two contextual capabilities $r_1 \times r_2$ is defined
  as the following  partial binary operation:
  \[
    r_1 \times r_2 \;\doteq\;
    \begin{cases}
      \{\, \sigma_1 \cup \sigma_2 \mid \sigma_1 \in r_1,\; \sigma_2 \in r_2 \,\}
        & \text{if } \mapCompat{r_1}{r_2}, \\
        \text{undefined}
        & \text{otherwise.}
      \end{cases}
  \]
\end{definition}

\begin{definition}[Contextual Capabilities Sum]\label{def:contextual-capabilities-sum}
  The sum of two contextual capabilities $r_1 + r_2$ is defined
  as the following partial binary operation:
  \[
    r_1 + r_2 \;\doteq\;
    \begin{cases}
      r_1 \cup r_2
        & \text{if } \Dom{r_1} = \Dom{r_2}, \\
        \text{undefined}
        & \text{otherwise.}
      \end{cases}
  \]
\end{definition}

Note that these two operations are \emph{partial}: two resources
cannot be multiplied unless they are compatible, and two resources
cannot be summed unless they have the same domain.

\begin{definition}[Contextual Capabilities Algebra]\label{def:contextual-capabilities-algebra}
  The tuple $(\Res,{+},{\times},\mathbf{1})$ forms a \emph{partial
    commutative semiring without zero} where the carrier set $\Res$ is
  the domain of all contextual capabilities, and the unit
  element $\mathbf{1}$ of product is the singleton capability
  $\{\emptyset\}$.
  The following algebraic properties hold:
  \begin{itemize}
  \item Identity: $\forall r \in \Res. r \times \mathbf{1} = \mathbf{1} \times r = r$.
  \item Commutative (when operations are defined):
  \begin{align*}
    \forall r_1\;r_2 \in \Res. r_1 + r_2 = r_2 + r_1 \\
    \forall r_1\;r_2 \in \Res. r_1 \times r_2 = r_2 \times r_1
  \end{align*}
  \item Associative (when operations are defined):
  \begin{align*}
    \forall r_1\;r_2\;r_3 \in \Res. (r_1 + r_2) + r_3 = r_1 + (r_2 + r_3) \\
    \forall r_1\;r_2\;r_3 \in \Res. (r_1 \times r_2) \times r_3 = r_1 \times (r_2 \times r_3)
  \end{align*}
  \end{itemize}
\end{definition}

\fi

\end{document}